\documentclass{midl} % Include author names

\usepackage{footnote}
\newcommand{\tabincell}[2]{\begin{tabular}{@{}#1@{}}#2\end{tabular}}
\usepackage{multirow}
\usepackage{booktabs}

\usepackage{mwe} % to get dummy images
\jmlrvolume{-- Under Review}
\jmlryear{2025}
\jmlrworkshop{Full Paper -- MIDL 2025 submission}
\editors{Under Review for MIDL 2025}

\title[Foundation model for Medical image class]{Rethinking Foundation Models for Medical Image Classification through a Benchmark Study on  MedMNIST}

\midlauthor{
\Name{Fuping Wu\nametag{$^{1,2}$}}
\thanks{Corresponding author} 
\Email{ fuping.wu@ndph.ox.ac.uk}\\
\addr $^{1}$ Nuffield Department of Population Health, University of Oxford, UK \\
\addr $^{2}$ Big Data Institute, University of Oxford, UK
\AND
\Name{Bart\l omiej W. Papie\.z\nametag{$^{1,2}$}} 
}

\begin{document}

\maketitle

\begin{abstract}
Foundation models are widely employed in medical image analysis, due to their high adaptability and generalizability for downstream tasks.
With the increasing number of foundation models being released, model selection has become an important issue.
In this work, we study the capabilities of foundation models in medical image classification tasks by conducting a benchmark study on the MedMNIST dataset. Specifically, we adopt various foundation models ranging from convolutional to Transformer-based models and implement both end-to-end training and linear probing for all classification tasks.
The results demonstrate the significant potential of these pre-trained models when transferred for medical image classification.
We further conduct experiments with different image sizes and various sizes of training data.
By analyzing all the results, we provide preliminary, yet useful insights and conclusions on this topic.
\end{abstract}


\section{Introduction}

Foundation models, which are pre-trained on large-scale diverse datasets, have achieved rapid advancement in various areas, such as computer vision \cite{awais2023foundational}, natural language processing \cite{zhou2023comprehensive}, and audio signal analysis \cite{huang2024audiogpt}.
These models have been proven effective in adapting to numerous downstream tasks, including semantic segmentation \cite{wang2024visionllm}, recommendation systems \cite{zhao2024recommender}, showcasing remarkable generalizability and adaptability \cite{touvron2023llama,anil2023palm}.

In recent years, the success of foundation models has triggered significant research into their application in medical image analysis \cite{zhang2024challenges}, either by transferring models from other fields \cite{mazurowski2023segment} or training models with large-scale medical data \cite{butoi2023universeg}.
While studies have shown the great potential of foundation models in organ segmentation \cite{MedSAM}, clinical reports generation \cite{thawakar2024xraygpt}, and medical Q$\&$A systems \cite{chen2024chexagent},
% While task-specific machine learning models are still the main methods used in clinical applications \cite{zhang2024challenges}, 
few have conducted large-scale comparisons of the existing models to provide insights into their performance in specific applications \cite{huix2024natural}.

In this work, we examine the performance of various foundation models in medical image classification tasks.
Although some studies have performed similar comparisons, they were often limited in scope, focusing on small model categories or datasets.
For example, \citet{huix2024natural} compared eight models across only four medical image classification tasks.
% , and showed that not all foundations transfer well to the medical domain.
\citet{wang2023real} presented a novel dataset with five classification tasks but compared only three models for transfer learning.
\citet{woerner2024navigating} conducted a benchmark study of few-shot and zero-shot medical image classification with linear probing solely focusing on fine-tuned ResNet-based models.
% , which made the comparison incomplete.
\citet{baharoon2023towards} compared DINOv2 \cite{oquab2023dinov2} with nine other models on four disease classification tasks.

To provide a comprehensive study of foundation models for medical image classification, we selected a diverse set of models, including four convolutional neural networks (CNN), and eight vision transformer (ViT)-based \cite{dosovitskiy2020image} models.
The ViT variants were trained for feature extraction or classification, offering deeper insights into their performance.
We used MedMNIST v2 \cite{yang2023medmnist}, a widely used benchmark for medical image classification, which comprises 12 distinct biomedical 2D datasets. 
The most relevant work to ours is by \citet{doerrich2024rethinking}, who also validated foundation models on MedMNIST v2 but used a different learning pipeline.
Notably, compared to \cite{doerrich2024rethinking}, we adopted more foundation models and achieved significantly better results for eight common models, leading to different conclusions.
This finding motivated us to conduct further analyses including the effect of image resizing techniques.

% compared 10 models on MedMNIST v2.
% Our work differs from that work in mainly three aspects:(1) We adopted more models including 8 common ones, facilitating a more comprehensive analysis; (2) We used different learning strategies for end-to-end training, and our results were significantly better than theirs, particularly for ViT-based models. This observation led to different conclusions and motivated us to embark on this work; (3) We analyzed other aspects of these models, such as image padding techniques.

The main contributions of this work are as follows.
\begin{itemize}
    \item We investigate various foundation models, demonstrating their potential in transferring to medical image classification tasks.

    \item We study the impact of image size and resizing strategies on the classification performance when transferring these foundation models, as well as their data efficiency.

    \item We analyze the results and provide several insights for transferring foundation models to medical image classification tasks, such as optimal learning rate settings for different model categories, model selection, and image resizing strategy.

\end{itemize}

\begin{figure}[t]
 % Caption and label go in the first argument and the figure contents
 % go in the second argument
\floatconts
  {fig:framework}
  {\caption{Framework of our study. We evaluate the performance using the MedMNIST dataset collection and select foundation models from a representative pool.}}
  {\includegraphics[width=0.9\linewidth]{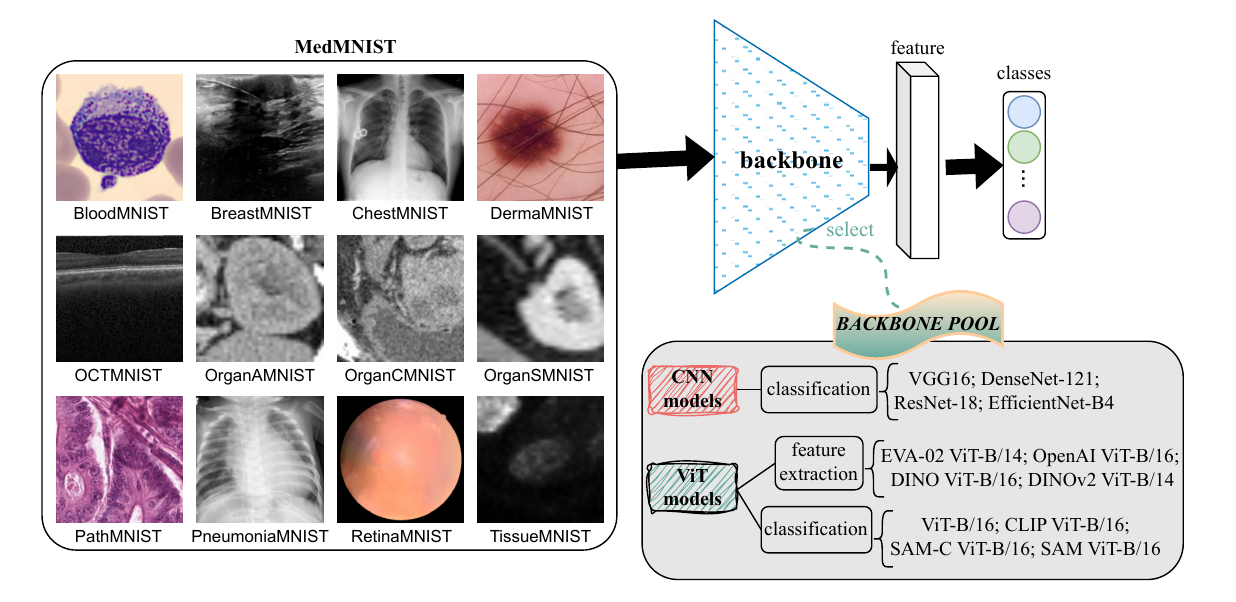}}
\end{figure}

\section{Method}

% Backbon_Dict = {'vgg16':'vgg16.tv_in1k',
%                 'resnet18':'resnet18.a1_in1k',
%                 'densenet121':'densenet121.ra_in1k',
%                 'effi_b4':'efficientnet_b4.ra2_in1k',
%                 'vit_b16':'vit_base_patch16_224.augreg2_in21k_ft_in1k',
%                 'clip_b16':'vit_base_patch16_clip_224.laion2b_ft_in12k_in1k',
                % 'eva_clip':  'eva02_base_patch14_224.mim_in22k', #    'eva02_base_patch16_clip_224.merged2b_s8b_b131k',
%                 'clip_opai':'vit_base_patch16_clip_224.openai_ft_in12k_in1k',

%                 'clip_opaif':'vit_base_patch16_clip_224.openai',
%                 'dino_base':'vit_base_patch16_224.dino',
%                 'dino_small':'vit_small_patch16_224.dino',
%                 'dino2_small':'vit_small_patch14_dinov2.lvd142m',
%                 'dino2_base':'vit_base_patch14_dinov2.lvd142m',
%                 'sam_base':'samvit_base_patch16.sa1b',
%                 'sam_cls':'vit_base_patch16_224.sam_in1k'}

\subsection{Pre-trained Model Selection}

With thousands of pre-trained models publicly available, we select representative backbones for validation and comparison, as shown in \figureref{fig:framework}, ranging from CNN-based to ViT-based architectures.
For CNN-based models, we include two baseline models: VGG16 \cite{simonyan2014very} and DenseNet-121 \cite{huang2017densely}, and two models with skip-connection via addition: EfficientNet-B4 \cite{tan2019efficientnet} and ResNet-18 \cite{he2016deep}.
All these CNN-based models were trained with ImageNet-1k \cite{russakovsky2015imagenet} for image classification.
For ViT-based models, we categorized them into two classes: 
(1) Four models trained for image classification: Three models, including ViT-B/16 \cite{steiner2021train}, CLIP ViT-B/16 \cite{ilharco_gabriel_2021_5143773} and SAM-C ViT-B/16 \cite{chen2021vision}, are adopted for comparison. These models were trained with different optimization strategies and datasets \cite{rw2019timm}, all including ImageNet-1k. We also included SAM ViT-B/16 \cite{kirillov2023segment} to compare with SAM-C ViT-B/16.
Although it was trained with SA-1B \footnote{\url{https://segment-anything.com}} for segmentation, its representations were demonstrated to be useful for various tasks.
(2) Four models trained for feature extraction: 
OpenAI ViT-B/16 \cite{ilharco_gabriel_2021_5143773} was trained on publicly available image-caption data to maximize the similarity of $(image, text)$ pairs; 
EVA-02 ViT-B/14 \cite{fang2024eva} can extract visual representation for reconstructing language-aligned vision features via masked image modeling;
DINO ViT-B/16 \cite{caron2021emerging} was trained with the self-supervised DINO method, and DINOv2 ViT-B/14 \cite{oquab2023dinov2} was trained with different losses and a larger dataset, namely LVD-142M.
% Finally, to observe the effect of model size, we involve in the smaller version of the two models, \textit{i.e.}, DINO ViT-S/16 and DINOv2 ViT-S/14.
% (3) Medical data-driven models.
% We choose MedSAM \cite{MedSAM} and BiomedGPT \cite{zhang2024generalist} for comparison.
% MedSAM is developed based on SAM for medical image segmentation and is trained on a large-scale medical image dataset.
% BiomedGPT ... 
For further details of these models, please refer to \tableref{tab:model details} in Appendix \ref{appendix:models}. 

\subsection{General Validation Framework }
As shown in~\figureref{fig:framework}, to validate the selected foundation models on different datasets for classification tasks, we append a one-layer linear classifier after the encoder of each model.
For a fair comparison, each classification model is trained with $15,000$ iterations, and optimized with the AdamW optimizer \cite{loshchilov2017decoupled} using two optimization strategies, linear probing and end-to-end fine-tuning.
In linear probing, the encoder is frozen and only the last classifier layer is optimized. The initial learning rate is set to $0.001$, which is reduced by 0.9 every 200 iterations.
For end-to-end fine-tuning, we adopt the same updating strategy as in linear probing for the classifier, while keeping the learning rate for the encoder, denoted as $lr_e$, as a constant.
Using a low learning rate for the encoder is a common method to adapt the pre-trained model to downstream tasks
\cite{baharoon2023towards}.
The selection of $lr_e$ will be discussed in Section \ref{sec:results}.

\section{Benchmark Results}
\subsection{Datasets}
To evaluate the potential of foundation models in medical image classification tasks, we utilize MedMNIST \cite{yang2023medmnist} collection for validation.
As illustrated in \figureref{fig:framework}, MedMNIST comprises 12 2D datasets, with modality spanning X-ray, ultrasound, CT, and electron microscope.
Each dataset is available in four different image resolutions: $28\times28$, $64\times64$, $128\times128$, and $224\times224$.
We primarily use images of $224\times224$ for testing.

During training, we employ the Cross-Entropy loss for multi-class classification and ordinal regression tasks, and the Binary Cross-Entropy loss for multi-label binary classification problems, such as the multi-label classification task for the ChestMNIST dataset.
% Details of the datasets can be found in the work of \citet{doerrich2024rethinking}.
To maintain consistency with \citet{yang2023medmnist},  all experiments are conducted three times on a single NVIDIA A100 GPU.
The performance of each model is assessed by the $mean \pm std$ of accuracy (ACC) and the area under the receiver
operating characteristic curve (AUC).

\subsection{Results}\label{sec:results}
\subsubsection{Learning rate selection for end-to-end fine-tuning strategy}
In the end-to-end fine-tuning process, we first determine a suitable learning rate, \textit{i.e.,} $lr_{e}$, for the encoder of each foundation model.
For the CNN models, we select $lr_{e}$ from $\{10^{-3},10^{-4},10^{-5}\}$, while for ViT-based models, we choose from $\{10^{-4},10^{-5},10^{-6}\}$. 
Figures \ref{fig:lr1} -\ref{fig:lr12} in Appendix \ref{appendix:tuning lr} illustrate the performance of the 12 foundation models across all 12 datasets with various learning rates.
% For the remaining 11 models, please refer to Figures \ref{fig:lr1} -\ref{fig:lr12} in Appendix for their performance changes with different $lr_{e}$.
Interestingly, the results indicate that for CNN models, $lr_{e}=10^{-4}$ generally yields optimal performance, while for ViT-based models, $lr_{e}=10^{-5}$ proves to be the best choice in most cases.

\subsubsection{End-to-end fine-tuning \textit{vs} linear probing}
\tableref{tab:results1} and Tables \ref{tab:results2}-\ref{tab:results4} in Appendix \ref{appendix:results_all} present the linear probing and the best end-to-end fine-tuning results for all foundation models across 12 datasets in the MedMNIST collection.
Although the 12 models were pre-trained with natural images or languages, they demonstrate significant potential in transfer learning for medical image classification. 
Notably, end-to-end fine-tuning consistently outperforms linear probing across all datasets, in both accuracy and AUC metrics, except the AUC value on the PathMNIST dataset, where the DINO ViT-B/16 achieves 99.78\% AUC with linear probing, 0.03\% higher than the former.
Particularly, on challenging tasks such as DermaMNIST, OCTMNIST, OrganCMNIST, OrganSMNIST, and TissueMNIST, end-to-end fine-tuning substantially outperforms linear probing.

\begin{savenotes}
\begin{table}[t]
\floatconts
  {tab:results1}%
  {\caption{Results of foundation models on DermaMNIST, PneumoniaMNIST, and BreastMNIST.}}%
  {\scalebox{0.9}{
  \begin{tabular}{lcccccc}
  \hline
   & \multicolumn{2}{c}{DermaMNIST}& \multicolumn{2}{c}{PneumoniaMNIST}& \multicolumn{2}{c}{BreastMNIST}\\
   \cmidrule(r){2-3}
   \cmidrule(r){4-5}
   \cmidrule(r){6-7}
   % \cmidrule(r){8-9}

   Methods &  ACC&  AUC&  ACC&  AUC& ACC &AUC \\
   \hline
   \multicolumn{7}{c}{END-TO-END}\\
   \hline
    VGG16& 85.49\tiny{$\pm$0.32}& 96.17\tiny{$\pm$0.55} & 90.12\tiny{$\pm$0.72} & 96.78\tiny{$\pm$1.15} & 90.60\tiny{$\pm$1.21} & 89.88\tiny{$\pm$1.02}  \\
    DenseNet-121    &  87.68\tiny{$\pm$0.75} & 96.49\tiny{$\pm$0.37} & \textbf{94.02\tiny{$\pm$0.59}} & 98.13\tiny{$\pm$0.89} & 91.24\tiny{$\pm$0.60} & 92.18\tiny{$\pm$0.73}  \\
    ResNet-18& 86.05\tiny{$\pm$0.78} & 96.62\tiny{$\pm$0.32} & 91.61\tiny{$\pm$0.79} & 97.90\tiny{$\pm$0.35} & 89.74\tiny{$\pm$0.91} & 92.47\tiny{$\pm$0.66}  \\
     EfficientNet-B4   & 88.13\tiny{$\pm$0.49} & 97.44\tiny{$\pm$0.07} & 90.38\tiny{$\pm$0.68} & 97.48\tiny{$\pm$0.03} & 87.61\tiny{$\pm$1.98} & 90.01\tiny{$\pm$1.38}  \\
     ViT-B/16   & 89.49\tiny{$\pm$0.24} & 98.25\tiny{$\pm$0.16} & 91.29\tiny{$\pm$0.72} & 97.81\tiny{$\pm$0.98} & \textbf{91.67\tiny{$\pm$0.52}} & \textbf{94.09\tiny{$\pm$0.89}} \\
     CLIP ViT-B/16   & 89.64\tiny{$\pm$0.31} & 98.31\tiny{$\pm$0.05} & 91.29\tiny{$\pm$0.42} & 98.80\tiny{$\pm$0.17} & 89.53\tiny{$\pm$1.21} & 91.02\tiny{$\pm$1.03} \\
     SAM-C ViT-B/16   & 88.43\tiny{$\pm$0.07} & 96.49\tiny{$\pm$0.41} & 91.24\tiny{$\pm$1.45} & 98.47\tiny{$\pm$0.16} & 90.60\tiny{$\pm$2.47} & 89.86\tiny{$\pm$1.34}  \\
     SAM ViT-B/16   & 80.62\tiny{$\pm$0.46} & 88.53\tiny{$\pm$1.30} & 88.25\tiny{$\pm$1.32} & 96.44\tiny{$\pm$0.29} & 86.75\tiny{$\pm$0.80} & 84.06\tiny{$\pm$0.91} \\
    EVA-02 ViT-B/14    & 88.79\tiny{$\pm$0.45} & 94.45\tiny{$\pm$1.27} & 90.76\tiny{$\pm$0.95} & 97.79\tiny{$\pm$0.38} & 87.61\tiny{$\pm$1.32} & 88.13\tiny{$\pm$0.55}  \\
      OpenAI ViT-B/16   &  89.74\tiny{$\pm$0.52} & 98.08\tiny{$\pm$0.27} & 90.01\tiny{$\pm$0.62} & 97.90\tiny{$\pm$0.33} & 89.96\tiny{$\pm$1.09} & 90.78\tiny{$\pm$0.82}  \\
    DINO ViT-B/16    & 89.39\tiny{$\pm$0.12} & 98.04\tiny{$\pm$0.21} & 91.56\tiny{$\pm$0.15} & 97.59\tiny{$\pm$1.16} & 88.46\tiny{$\pm$1.05} & 91.38\tiny{$\pm$0.64}  \\
    % DINO ViT-S/16    & 88.46\tiny{$\pm$0.49} & 98.12\tiny{$\pm$0.14} & 91.35\tiny{$\pm$0.80} & 98.47\tiny{$\pm$1.16} & 88.68\tiny{$\pm$0.80} & 89.71\tiny{$\pm$1.03}  \\
    DINOv2 ViT-B/14    &\textbf{ 92.02\tiny{$\pm$0.32}} & \textbf{98.92\tiny{$\pm$0.11}} & 90.81\tiny{$\pm$0.20} & \textbf{99.24\tiny{$\pm$0.14}} & 90.81\tiny{$\pm$1.68} & 92.00\tiny{$\pm$1.99}  \\
    % DINOv2 ViT-S/14    & 90.76\tiny{$\pm$0.31} & 98.51\tiny{$\pm$0.03} & 89.96\tiny{$\pm$1.91} & 98.43\tiny{$\pm$0.53} & 89.96\tiny{$\pm$0.80} & 91.50\tiny{$\pm$0.98} \\
    % MedSAM ViT-B/16     & \tiny{$\pm$} & \tiny{$\pm$} & \tiny{$\pm$} & \tiny{$\pm$} & \tiny{$\pm$} & \tiny{$\pm$} \\
    %  BiomedGPT  &  \tiny{$\pm$} & \tiny{$\pm$} & \tiny{$\pm$} & \tiny{$\pm$} & \tiny{$\pm$} & \tiny{$\pm$}  \\
           \hline
   \multicolumn{7}{c}{LINEAR PROBING}\\
   \hline
    VGG16& 76.60\tiny{$\pm$0.21}& 92.28\tiny{$\pm$0.01} & 86.59\tiny{$\pm$0.19} & 97.23\tiny{$\pm$0.02} & 84.18\tiny{$\pm$0.79} & 88.61\tiny{$\pm$0.24}  \\
    DenseNet-121    &  78.00\tiny{$\pm$0.34} & 93.14\tiny{$\pm$0.03} & 87.39\tiny{$\pm$0.45} & 97.28\tiny{$\pm$0.01} & 82.47\tiny{$\pm$0.79} & 81.15\tiny{$\pm$0.27}  \\
    ResNet-18& 74.79\tiny{$\pm$0.10} & 90.87\tiny{$\pm$0.01} & 85.25\tiny{$\pm$0.59} & 95.25\tiny{$\pm$0.06} & 81.83\tiny{$\pm$0.30} & 81.53\tiny{$\pm$0.21}  \\
     EfficientNet-B4   & 72.38\tiny{$\pm$0.55} & 89.28\tiny{$\pm$0.42} & 85.52\tiny{$\pm$0.42} & 94.10\tiny{$\pm$0.53} & 75.42\tiny{$\pm$1.98} & 77.40\tiny{$\pm$1.85}  \\
     ViT-B/16   & 83.40\tiny{$\pm$0.26} & 96.45\tiny{$\pm$0.01} & 88.30\tiny{$\pm$0.81} & 97.85\tiny{$\pm$0.13} & \textbf{89.52\tiny{$\pm$0.60}} & \textbf{93.74\tiny{$\pm$0.17}}  \\
     CLIP ViT-B/16   & 82.36\tiny{$\pm$0.16} & 96.11\tiny{$\pm$0.05} & 88.24\tiny{$\pm$0.15} & 97.81\tiny{$\pm$0.04} & 85.25\tiny{$\pm$0.52} & 89.91\tiny{$\pm$0.19}  \\
     SAM-C ViT-B/16   & 83.04\tiny{$\pm$0.08} & 96.49\tiny{$\pm$0.02} & 89.36\tiny{$\pm$0.15} & 98.46\tiny{$\pm$0.02} & 87.39\tiny{$\pm$0.30} & 92.44\tiny{$\pm$0.07}\\
     SAM ViT-B/16   & 66.88\tiny{$\pm$0.00} & 51.00\tiny{$\pm$0.53} & 73.07\tiny{$\pm$0.26} & 90.57\tiny{$\pm$0.04} & 76.49\tiny{$\pm$0.30} & 76.42\tiny{$\pm$0.02}  \\
    EVA-02 ViT-B/14    & 82.70\tiny{$\pm$0.20} & 96.09\tiny{$\pm$0.02} & 86.21\tiny{$\pm$0.00} & 96.82\tiny{$\pm$0.02} & 87.82\tiny{$\pm$0.00} & 90.84\tiny{$\pm$0.05}  \\
      OpenAI ViT-B/16   &  82.80\tiny{$\pm$0.23} & 96.14\tiny{$\pm$0.04} & 86.91\tiny{$\pm$0.15} & 96.75\tiny{$\pm$0.00} & 86.75\tiny{$\pm$0.30} & 90.64\tiny{$\pm$0.28}  \\
    DINO ViT-B/16    & \textbf{84.87\tiny{$\pm$0.36}} & \textbf{97.06\tiny{$\pm$0.02}} & \textbf{92.41\tiny{$\pm$0.42}} & \textbf{98.87\tiny{$\pm$0.03}} & 88.24\tiny{$\pm$0.30} & 92.73\tiny{$\pm$0.07} \\
    % DINO ViT-S/16    & 84.03\tiny{$\pm$0.24} & 96.25\tiny{$\pm$0.08} & 93.85\tiny{$\pm$0.15} & 99.29\tiny{$\pm$0.02} & 86.53\tiny{$\pm$0.90} & 93.49\tiny{$\pm$0.12}  \\
    DINOv2 ViT-B/14    & 83.97\tiny{$\pm$0.08} & 96.62\tiny{$\pm$0.02} & 90.65\tiny{$\pm$0.65} & 98.41\tiny{$\pm$0.08} & 87.39\tiny{$\pm$1.08} & 92.50\tiny{$\pm$0.04}  \\
    % DINOv2 ViT-S/14    & 82.52\tiny{$\pm$0.28} & 95.99\tiny{$\pm$0.02} & 88.24\tiny{$\pm$0.67} & 98.19\tiny{$\pm$0.06} & 87.39\tiny{$\pm$0.30} & 93.03\tiny{$\pm$0.27}  \\
    % MedSAM ViT-B/16     & \tiny{$\pm$} & \tiny{$\pm$} & \tiny{$\pm$} & \tiny{$\pm$} & \tiny{$\pm$} & \tiny{$\pm$}  \\
    %  BiomedGPT  &  \tiny{$\pm$} & \tiny{$\pm$} & \tiny{$\pm$} & \tiny{$\pm$} & \tiny{$\pm$} & \tiny{$\pm$} \\
           \hline
              \multicolumn{7}{c}{Results from \citet{zhang2024generalist} for models pre-trained on medical data}\\
   \hline
BiomedGPT    & 86.6 &-  &-  &-  &79.5  &-  \\
BiomedCLIP    & 71.9 &-  & - & - &82.2  &-  \\
MedSAM    & 72.3 & - & - &-  &76.9  & - \\

   \hline
  \end{tabular}}
  }
\end{table}
\end{savenotes}

\subsubsection{CNN models \textit{vs} ViT-based models}
When comparing the results of CNN models with those of ViT-based models, we observe that most of the best performance is achieved by ViT-based models in both linear probing and end-to-end fine-tuning settings. 
Notably, DenseNet-121 and EfficientNet-B4 achieve the highest accuracy on PneumoniaMNIST and the highest AUC on BloodMNIST, but only under the end-to-end fine-tuning strategy.

Among ViT-based methods, ViT-B/16 generally excels in the category of classification-targeted pre-trained models. For feature extraction models, DINO or DINOv2 frequently delivers the best results.
Interestingly, SAM-C ViT-B/16 consistently outperforms SAM ViT-B/16 across all classification tasks, especially with the linear probing strategy.
This may be attributed to SAM being primarily trained for segmentation tasks, which could hinder its ability to capture global information effectively for classification purposes.
%The reason can be that the latter was trained for segmentation tasks, thus could be difficult to capture the global information.

\subsubsection{Impact of training pipeline}
It is important to note that \citet{doerrich2024rethinking} adopted a different training pipeline when validating foundation models on MedMNIST, which led to conclusions that diverged from ours.
Specifically, they employed the AdamW optimizer with a learning rate of 0.0001 and a cosine annealing learning rate scheduler \cite{loshchilov2022sgdr} with a single cycle.
When comparing the results for the 8 common models, highlighted in \tableref{tab:model details} in Appendix \ref{appendix:models}, we found similar outcomes under the linear probing strategy.
However, our results surpass theirs in the end-to-end fine-tuning setting, especially for ViT-based models.
For instance, on BreastMNIST, ViT-B/16 achieved an accuracy of $83.76\pm 1.09$ and an AUC of $86.18\pm 0.26$ of AUC in their setup. In contrast, our setup yielded better results, with an accuracy of $91.67\pm 0.52$ and an AUC of $94.09\pm 0.89$.
Similarly, on DermaMNIST, DINO ViT-B/16 achieved an accuracy of $81.31\pm 1.05$ in their work \textit{vs} $89.39\pm 0.12$ in ours.

Based on a different training pipeline, \citet{doerrich2024rethinking} delivered several conclusions that could be contradictory to the results in this work.
For example, they concluded that self-supervised pretraining models, such as CLIP and DINO, do not always improve medical image classification results in end-to-end fine-tuning while demonstrating enhanced performance with linear probing.
Conversely, our results demonstrate that these models not only improve the performance in end-to-end fine-tuning, but also outperform linear probing.
Additionally, they concluded that CNN models consistently outperform ViT-based models in accuracy with end-to-end training, while in our work we have an almost opposite conclusion.
For example, on the DermaMNIST dataset, DINOv2 ViT-B/14, reached the highest accuracy of $92.02\pm 0.32$, while the best-performing CNN model, EfficientNet-B4, obtained an accuracy of $88.13\pm 0.49$.

\section{Discussion}

\begin{savenotes}
\begin{table}[t]
\floatconts
  {tab:results_sizes_scale}%
  {\caption{Results of foundation models on DermaMNIST of different image sizes using scaling to resize the original images into a resolution of $224\times 224$.}}%
  {\scalebox{0.9}{
  \begin{tabular}{lcccccc}
  \hline
   & \multicolumn{2}{c}{$28\times 28$}& \multicolumn{2}{c}{$64\times 64$}& \multicolumn{2}{c}{$128\times 128$}\\
   \cmidrule(r){2-3}
   \cmidrule(r){4-5}
   \cmidrule(r){6-7}
   % \cmidrule(r){8-9}

   Methods &  ACC&  AUC&  ACC&  AUC& ACC &AUC \\
   \hline
   \multicolumn{7}{c}{END-TO-END}\\
   \hline
    VGG16 & 77.36\tiny{$\pm$0.32} & 91.76\tiny{$\pm$0.40}  &83.62\tiny{$\pm$0.20}   &95.91\tiny{$\pm$0.28}   &85.35\tiny{$\pm$0.80}   &96.10\tiny{$\pm$0.23}  \\
    DenseNet-121    & 78.04\tiny{$\pm$0.51} & 90.29\tiny{$\pm$1.03}  &85.15\tiny{$\pm$0.08}   &95.51\tiny{$\pm$0.47}   &86.57\tiny{$\pm$0.65}   &96.43\tiny{$\pm$0.50}  \\
    ResNet-18 &79.83\tiny{$\pm$0.58}  &92.98\tiny{$\pm$0.28}   &84.80\tiny{$\pm$1.09}   &95.74\tiny{$\pm$0.44}   &85.80\tiny{$\pm$0.68}   & 96.46\tiny{$\pm$0.38} \\
     EfficientNet-B4  & 76.31\tiny{$\pm$1.11} & 91.37\tiny{$\pm$0.46}  & 84.42\tiny{$\pm$0.80}  &95.66\tiny{$\pm$0.43}   &86.37\tiny{$\pm$0.94}   &96.75\tiny{$\pm$0.31}  \\
     ViT-B/16    &81.68\tiny{$\pm$0.96}  &95.00\tiny{$\pm$0.21}   &87.65\tiny{$\pm$0.21}   &97.80\tiny{$\pm$0.24}   & 88.66\tiny{$\pm$0.10}  &98.02\tiny{$\pm$0.15}  \\
     CLIP ViT-B/16  & 81.60\tiny{$\pm$0.40} &94.63\tiny{$\pm$0.15}   &87.96\tiny{$\pm$0.60}   &97.67\tiny{$\pm$0.10}   &89.21\tiny{$\pm$0.16}   &98.13\tiny{$\pm$0.04}  \\
     SAM-C ViT-B/16   &81.55\tiny{$\pm$0.32}  & 92.87\tiny{$\pm$0.82}  & 87.25\tiny{$\pm$0.38}  &96.45\tiny{$\pm$0.17}   &87.48\tiny{$\pm$0.84}   &96.89\tiny{$\pm$0.10}  \\
     SAM ViT-B/16   &77.01\tiny{$\pm$1.21}  & 86.51\tiny{$\pm$1.44}  &80.02\tiny{$\pm$0.33}   &88.23\tiny{$\pm$1.35}   & 79.98\tiny{$\pm$1.14}  &87.89\tiny{$\pm$0.41}  \\
    EVA-02 ViT-B/14    &80.35\tiny{$\pm$0.36}  & 84.75\tiny{$\pm$0.57}  & 86.78\tiny{$\pm$0.41}  &92.26\tiny{$\pm$0.71}   &87.65\tiny{$\pm$0.44}   &92.63\tiny{$\pm$0.67}  \\
      OpenAI ViT-B/16    &81.31\tiny{$\pm$0.24}  &94.43\tiny{$\pm$0.30}   &88.26\tiny{$\pm$0.72}   &97.94\tiny{$\pm$0.18}   &89.33\tiny{$\pm$0.12}   & 98.00\tiny{$\pm$0.05} \\
    DINO ViT-B/16     & 80.53\tiny{$\pm$0.40} &94.16\tiny{$\pm$0.45}   & 86.75\tiny{$\pm$0.28}  & 97.47\tiny{$\pm$0.13}  & 88.10\tiny{$\pm$0.45}  &97.72\tiny{$\pm$0.06}  \\
    % DINO ViT-S/16    &  &   &   &   &   &  \\
    DINOv2 ViT-B/14  & \textbf{82.19\tiny{$\pm$0.25}} & \textbf{95.30\tiny{$\pm$0.10}}  & \textbf{89.83\tiny{$\pm$0.25}}  &\textbf{98.41\tiny{$\pm$0.10}}   & \textbf{91.19\tiny{$\pm$0.48}}  & \textbf{98.63\tiny{$\pm$0.01}} \\
    % DINOv2 ViT-S/14    &  &   &   &   &   &  \\
    % MedSAM ViT-B/16     &  &   &   &   &   &  \\
    %  BiomedGPT   &  &   &   &   &   &  \\
           \hline
   \multicolumn{7}{c}{LINEAR PROBING}\\
   \hline
    VGG16 & 74.96\tiny{$\pm$0.11} &90.67\tiny{$\pm$0.03}   &77.04\tiny{$\pm$0.29}   &92.38\tiny{$\pm$0.02}   &76.48\tiny{$\pm$0.37}   &92.49\tiny{$\pm$0.12}  \\
    DenseNet-121    & 73.23\tiny{$\pm$0.16} &90.69\tiny{$\pm$0.14}   &77.04\tiny{$\pm$0.43}   & 92.66\tiny{$\pm$0.05}  &76.94\tiny{$\pm$0.29}   & 93.01\tiny{$\pm$0.05} \\
    ResNet-18 & 73.18\tiny{$\pm$0.13} & 87.69\tiny{$\pm$0.07}  &75.01\tiny{$\pm$0.50}   &90.73\tiny{$\pm$0.11}   &75.10\tiny{$\pm$0.06}  &91.47\tiny{$\pm$0.06}  \\
     EfficientNet-B4  & 69.38\tiny{$\pm$0.46} &84.40\tiny{$\pm$0.65}   &72.85\tiny{$\pm$0.51}   &88.64\tiny{$\pm$0.20}   &72.40\tiny{$\pm$0.55}   &88.91\tiny{$\pm$0.59}  \\
     ViT-B/16    & 77.27\tiny{$\pm$0.35} & 93.23\tiny{$\pm$0.05}  & 81.23\tiny{$\pm$0.29}  &95.36\tiny{$\pm$0.08}   & 82.61\tiny{$\pm$0.52}  &96.12\tiny{$\pm$0.16}  \\
     CLIP ViT-B/16  & 75.11\tiny{$\pm$0.22} &92.84\tiny{$\pm$0.12}   &79.87\tiny{$\pm$0.22}   &94.81\tiny{$\pm$0.04}   & 80.76\tiny{$\pm$0.19}  & 95.45\tiny{$\pm$0.05} \\
     SAM-C ViT-B/16   & 77.37\tiny{$\pm$0.23} &93.92\tiny{$\pm$0.09}   & 80.63\tiny{$\pm$0.72}  &95.63\tiny{$\pm$0.23}   & 81.58\tiny{$\pm$0.25}  &96.10\tiny{$\pm$0.06}  \\
     SAM ViT-B/16   &66.88\tiny{$\pm$0.00}  & 43.72\tiny{$\pm$0.24}  & 66.88\tiny{$\pm$0.00}  &45.91\tiny{$\pm$0.34}   & 66.88\tiny{$\pm$0.00}  &49.15\tiny{$\pm$0.54}  \\
    EVA-02 ViT-B/14    &78.09\tiny{$\pm$0.02}  & \textbf{94.13\tiny{$\pm$0.01}}  &82.76\tiny{$\pm$0.13}   &96.12\tiny{$\pm$0.00}   &82.34\tiny{$\pm$0.18}   &95.82\tiny{$\pm$0.03}  \\
      OpenAI ViT-B/16    &76.77\tiny{$\pm$0.25}  &93.03\tiny{$\pm$0.09}   & 80.07\tiny{$\pm$0.19}  & 95.19\tiny{$\pm$0.01}  & 82.59\tiny{$\pm$0.16}  & 95.81\tiny{$\pm$0.01} \\
    DINO ViT-B/16     &\textbf{78.30\tiny{$\pm$0.41}}  & 93.47\tiny{$\pm$0.03}  &\textbf{82.78\tiny{$\pm$0.12}}   &\textbf{96.20\tiny{$\pm$0.04}}   & \textbf{84.56\tiny{$\pm$0.26}}  &\textbf{96.79\tiny{$\pm$0.01}}  \\
    % DINO ViT-S/16    &  &   &   &   &   &  \\
    DINOv2 ViT-B/14  & 75.56\tiny{$\pm$0.27} &91.75\tiny{$\pm$0.08}   &80.45\tiny{$\pm$0.25}   &94.98\tiny{$\pm$0.02}   & 82.61\tiny{$\pm$0.34}  & 96.09\tiny{$\pm$0.08} \\
    % DINOv2 ViT-S/14    &  &   &   &   &   &  \\
    % MedSAM ViT-B/16     &  &   &   &   &   &  \\
    %  BiomedGPT   &  &   &   &   &   &  \\
           \hline
  \end{tabular}}
  }
\end{table}
\end{savenotes}
\subsection{Effect of Image Size and Resizing Strategy}
%We further studied the effect of resizing strategies when the image size is small, as many foundation models require a fixed input size.
We further investigated the impact of resizing strategies on performance, particularly when handling smaller image sizes, as many foundation models require a fixed input size. 
Specifically, we compare zero-padding and scaling strategies for both end-to-end fine-tuning and linear probing, focusing on the DermaMNIST dataset with image sizes of $28\times 28$, $64\times 64$ and $128\times 128$.
\tableref{tab:results_sizes_scale} and \tableref{tab:results_sizes_pad} in Appendix \ref{appendix:results_derma_sizes} present the results with scaling and zero-padding strategies, respectively.
The findings reveal that, generally, accuracy improves with larger image sizes when using zero-padding. 
Similarly, for the scaling strategy, larger image sizes typically result in higher accuracy. However, the differences, particularly among sizes of $64\times 64$, $128\times 128$, and $224\times 224$  are marginal.
Comparison plots illustrating these trends can be found in Figures \ref{fig:ft_scale_sizes}-\ref{fig:fz_pad_sizes} in Appendix \ref{appendix:results_plot_resize}.
Notably, for end-to-end fine-tuning, scaling tends to yield better accuracy than zero-padding across all image sizes.
Conversely, for linear probing, the performance gap between scaling and zero-padding narrows, particularly when the image size is $128\times 128$.
For a detailed comparison, please refer to Figures \ref{fig:ft_28}-\ref{fig:fz_128} in Appendix \ref{appendix:results_plot_resize}.

\subsection{Data Efficiency}

\begin{figure}[t]
 % Caption and label go in the first argument and the figure contents
 % go in the second argument
\floatconts
  {fig:few1}
  {\caption{Accuracy of DINO ViT-B/16 on DermaMNIST with various numbers of training data for each class, and ``full'' means using all training data.}}
  {
\includegraphics[width=1.0\linewidth]{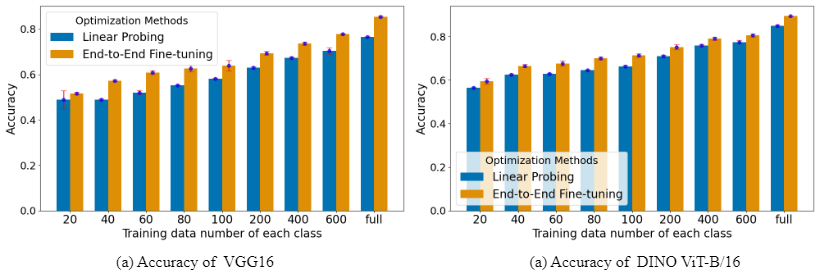}%
}
\end{figure}

We further investigated whether these foundation models can be adapted to medical image classification tasks with limited training data using end-to-end fine-tuning or linear probing.
This analysis was conducted on the DermaMNIST dataset, varying the number of training images per class in $\{20,40,60,80,100,200,400,600\}$.
As illustrated in \figureref{fig:few1}, the accuracy of DINO ViT-B/16 improves significantly with an increasing number of training images for both end-to-end fine-tuning and linear probing.
This indicates the low data efficiency of DINO ViT-B/16 when transferring to this task.
Similar trends were observed across the other 11 models as detailed in Figures \ref{fig:few3}-\ref{fig:few12} in Appendix \ref{appendix:plots_few_shot}.
These findings suggest that few-shot learning for foundation model transferring remains challenging, and requires delicate optimization strategies.

\subsection{Transferring Foundation Models Pre-trained on Medical Data}

\citet{zhang2024generalist} fine-tuned three foundation models pre-trained on medical data, including BioMedGPT \cite{zhang2024generalist}, BioMedCLIP \citet{zhang2023biomedclip} and MedSAM \cite{MedSAM}, on seven datasets from the MedMNIST collection.
Their results, presented at the bottom of \tableref{tab:results1} and Tables \ref{tab:results2}-\ref{tab:results4} in Appendix \ref{appendix:results_all} are generally poorer compared to the models we adopted.
%This suggests more efforts for developing medical data-based foundation models.
%However, the results of linear probing are missing for these models, making the comparison incomplete, we leave it as a future work.
This suggests the need for further development of foundation models based on medical data. However, the absence of linear probing results for these models limits the completeness of the comparison, which we propose to address in future work.
% \subsection{Limitations}

\section{Conclusion}

In this work, we evaluated the performance of foundation models on medical image classification tasks, selecting 12 representative models spanning CNN and ViT-based models. We employed both end-to-end fine-tuning and linear probing strategies for model transfer.
Our results demonstrate the significant potential of foundation models in these tasks.
Additionally, we explored the impact of image size and resizing strategies, deriving useful insights from our findings.
Despite these promising results, several limitations remain: 
(1) Our validation is limited to the MedMNIST collection. Including additional datasets would provide more comprehensive conclusions;
% , particularly real-world clinical data, such as images in UK BioBank
(2) Our analysis of resizing strategies is limited to the DermaMNIST dataset. A broader validation across all 12 datasets in the collection is planned for future work. 
%For MedMNIST collection, we solely used DermaMNIST dataset when studying the effect of the image resizing strategy. A complete validation on all 12 datasets will be implemented in the future; 
(3)
The current analysis lacks depth, particularly regarding model performance on specific classes within each task. A more detailed discussion will be provided.
% The analysis of the results is superficial, we will discuss them with more details, such as for each task, how models perform for each specific class; and 
(4) The biases of each model across different medical image classification tasks are yet to be understood \cite{alloula2024biases}.
% the biases of each model on different medical image classification tasks remain unknown.

\clearpage  % Acknowledgements, references, and appendix do not count toward the page limit (if any)
% Acknowledgments---Will not appear in anonymized version
\midlacknowledgments{The authors would like to acknowledge the funding from Novo Nordisk to support this work.}

\bibliography{midl-samplebibliography}

% \newpage
\clearpage
\appendix

\section{Details of the Evaluated Models.} \label{appendix:models}

\begin{savenotes}
\begin{table}[h]
\floatconts
  {tab:model details}%
  {\caption{Details of the 12 evaluated models. `\checkmark' means the model was also validated in the work of \citet{doerrich2024rethinking}.}}%
  {\scalebox{0.85}{
  \begin{tabular}{|l|c|c|c|c|}
  \hline
  \bfseries Model & \bfseries \tabincell{l}{Params\\ \quad (M)}  & \bfseries \tabincell{l}{Model\\ \ Task} & \bfseries \tabincell{l}{Training\\ \quad Data} & \bfseries Source\\
  \hline
   \multicolumn{5}{c}{CNN-based Models} \\
   \hline
   VGG16 (\checkmark) & $138.4$  & \tabincell{l}{\quad Image\\ classification} & ImageNet-1k & \tabincell{l}{timm\footnote{
timm at Hugginface: \url{https://huggingface.co/timm}} with identifier:\\ \textit{vgg16.tv\_in1k}}\\
   \hline
   ResNet-18 (\checkmark)& 11.7 &\tabincell{l}{\quad Image\\ classification} &ImageNet-1k & \tabincell{l}{timm with identifier:\\ \textit{resnet18.a1\_in1k}}\\
      \hline
   DenseNet121 (\checkmark)&8.0 &\tabincell{l}{\quad Image\\ classification} &ImageNet-1k &\tabincell{l}{timm with identifier:\\ \textit{densenet121.ra\_in1k}}  \\
   \hline
      EfficientNet-B4 (\checkmark) &19.3 &\tabincell{l}{\quad Image\\ classification} &ImageNet-1k &\tabincell{l}{timm with identifier:\\ \textit{efficientnet\_b4.ra2\_in1k}} \\
      \hline
         \multicolumn{5}{c}{Transformer-based Models} \\
   \hline
         ViT-B/16 (\checkmark)&86.6 &\tabincell{l}{\quad Image\\ classification} &\tabincell{l}{ImageNet-21k,\\ImageNet-1k} &\tabincell{l}{timm with identifier:\\ \textit{vit\_base\_patch16\_224.}\\\textit{augreg2\_in21k\_ft\_in1k}} \\
      \hline

           CLIP ViT-B/16 (\checkmark)    &86.6 &\tabincell{l}{\quad Image\\ classification} &\tabincell{l}{LAION-2B, \\ImageNet-12k,\\ImageNet-1k} &\tabincell{l}{timm with identifier:\\ \textit{vit\_base\_patch16\_clip\_224.}\\\textit{laion2b\_ft\_in12k\_in1k}} \\
                 \hline
              SAM-C ViT-B/16 &86.6 &\tabincell{l}{\quad Image\\ classification} &ImageNet-1k &\tabincell{l}{timm with identifier:\\ \textit{vit\_base\_patch16\_224.}\\\textit{sam\_in1k}} \\
           \hline
                    SAM ViT-B/16 (\checkmark) &89.7 &\tabincell{l}{ Image\\ segmentation} &SA-1B &\tabincell{l}{timm with identifier:\\ \textit{samvit\_base\_patch16.}\\\textit{sa1b}} \\

      \hline
      
            EVA-02 ViT-B/14   &85.8  &\tabincell{l}{Feature\\ extraction} &ImageNet-22k &\tabincell{l}{timm with identifier:\\ \textit{eva02\_base\_patch14\_224.}\\\textit{mim\_in22k}} \\
      \hline
             OpenAI ViT-B/16  & 86.6 &\tabincell{l}{ Feature\\ extraction} &\tabincell{l}{ Image-text \\ open source} & \tabincell{l}{timm with identifier:\\ \textit{vit\_base\_patch16\_clip\_224.}\\\textit{openai}}\\
      \hline
              DINO ViT-B/16 (\checkmark) &85.8 &\tabincell{l}{ Feature\\ extraction} &ImageNet-1k &\tabincell{l}{timm with identifier:\\ \textit{vit\_base\_patch16\_224.}\\\textit{dino}} \\
      % \hline
      %         DINO ViT-S/16 &21.7 & &\tabincell{l}{ Feature\\ extraction} &ImageNet-1k &\tabincell{l}{timm with identifier:\\ \textit{vit\_small\_patch16\_224.}\\\textit{dino}}  \\
      \hline
               DINOv2 ViT-B/14&86.6 &\tabincell{l}{ Feature\\ extraction} &LVD-142M &\tabincell{l}{timm with identifier:\\ \textit{vit\_base\_patch14\_dinov2.}\\\textit{lvd142m}} \\
      % \hline
      %         DINOv2 ViT-S/14 &22.1 & &\tabincell{l}{ Feature\\ extraction} &LVD-142M &\tabincell{l}{timm with identifier:\\ \textit{vit\_small\_patch14\_dinov2.}\\\textit{lvd142m}} \\
      
      \hline

              MedSAM ViT-B/16&89.7 &\tabincell{l}{\quad Image\\ segmentation} &\tabincell{l}{ ImageNet-1k, \\ Medical image \\open source} & Github repository \footnote{\url{https://github.com/bowang-lab/MedSAM}} \\
      \hline
  \end{tabular}}
  }
\end{table}
\end{savenotes}

\section{Tuning Learning Rate for Foundation Models}\label{appendix:tuning lr}
\begin{figure}[h]
 % Caption and label go in the first argument and the figure contents
 % go in the second argument
\floatconts
  {fig:lr1}
  {\caption{
  % Tuning the learning rate for the encoder of VGG16 ranging in $\{10^{-3},10^{-4},10^{-5}\}$.
  Comparing the Accuracy of VGG16 with the learning rate of the encoder ranging in $\{10^{-3},10^{-4},10^{-5}\}$ on different datasets.}}
  {\includegraphics[width=0.7\linewidth]{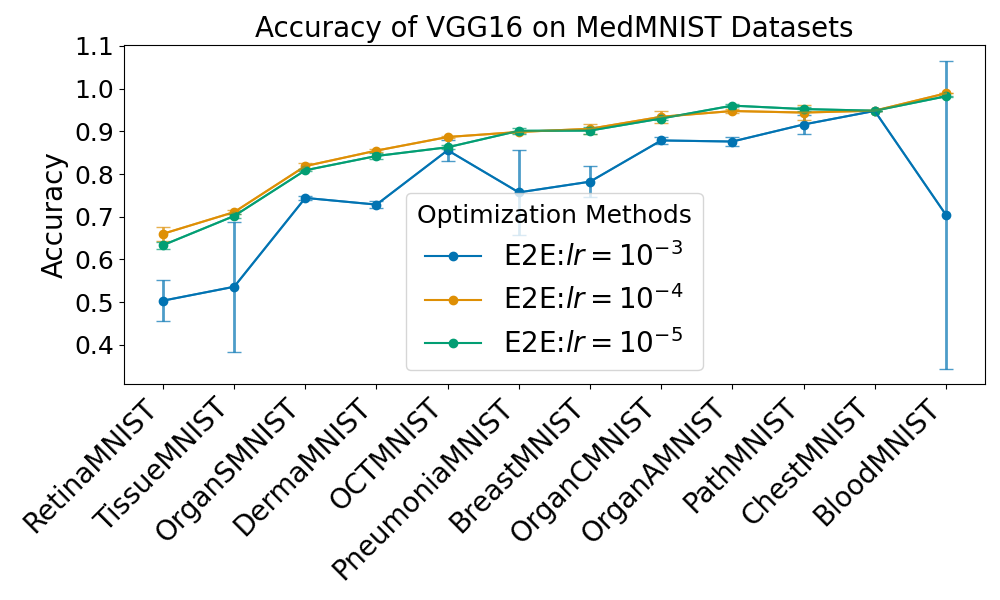}}%

\end{figure}

\begin{figure}[h]
 % Caption and label go in the first argument and the figure contents
 % go in the second argument
\floatconts
  {fig:lr3}
  {\caption{Comparing the Accuracy of ResNet-18 with the learning rate of the encoder ranging in $\{10^{-3},10^{-4},10^{-5}\}$ on different datasets.}}
  {
\includegraphics[width=0.7\linewidth]{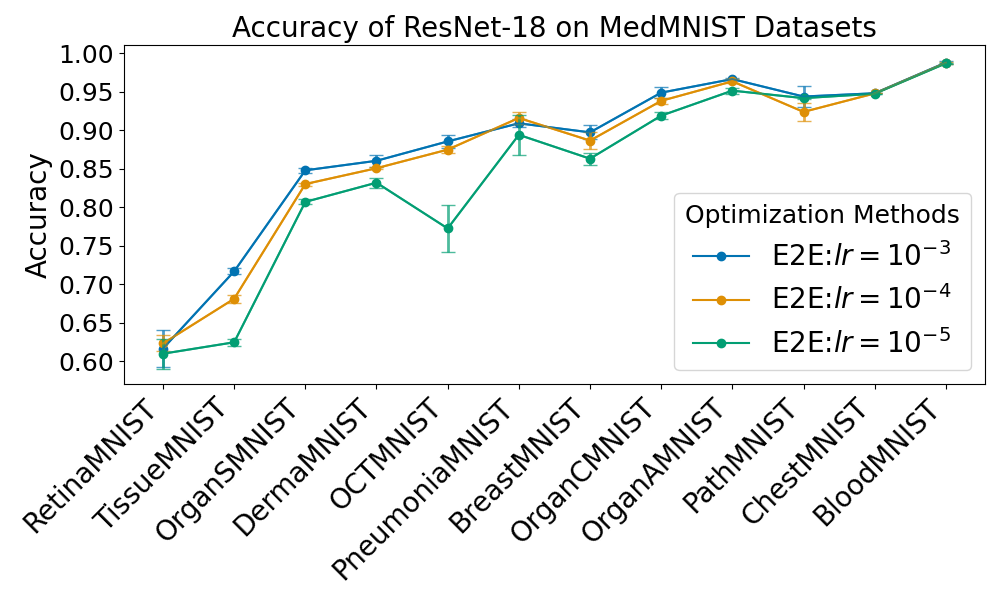}%
}
\end{figure}

\begin{figure}[h]
 % Caption and label go in the first argument and the figure contents
 % go in the second argument
\floatconts
  {fig:lr4}
  {\caption{Comparing the Accuracy of DenseNet-121 with the learning rate of the encoder ranging in $\{10^{-3},10^{-4},10^{-5}\}$ on different datasets.}}
  {
\includegraphics[width=0.7\linewidth]{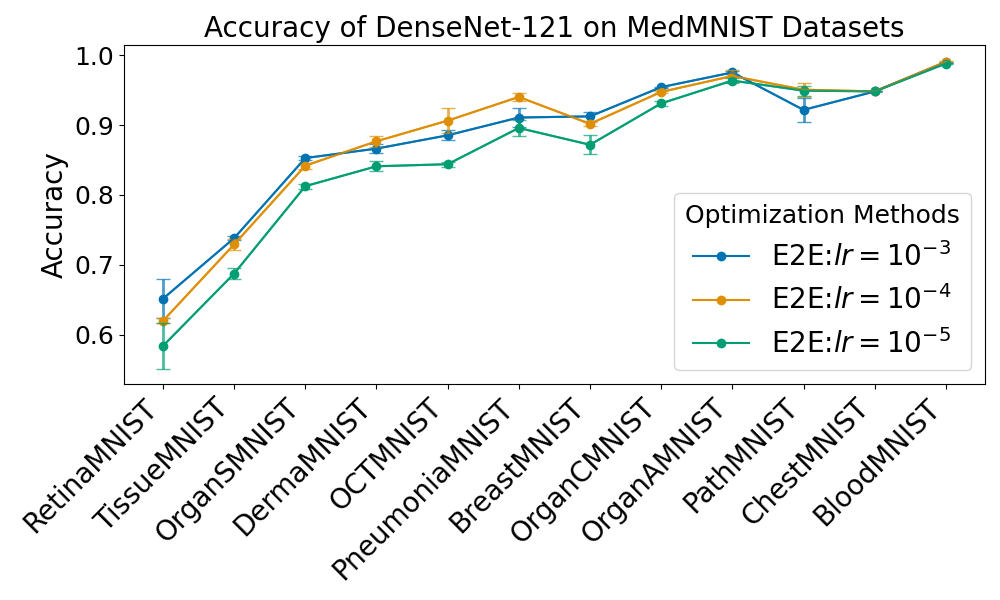}%
}
\end{figure}

\begin{figure}[h]
 % Caption and label go in the first argument and the figure contents
 % go in the second argument
\floatconts
  {fig:lr5}
  {\caption{Comparing the Accuracy of EfficientNet-B4 with the learning rate of the encoder ranging in $\{10^{-3},10^{-4},10^{-5}\}$ on different datasets.}}
  {
\includegraphics[width=0.7\linewidth]{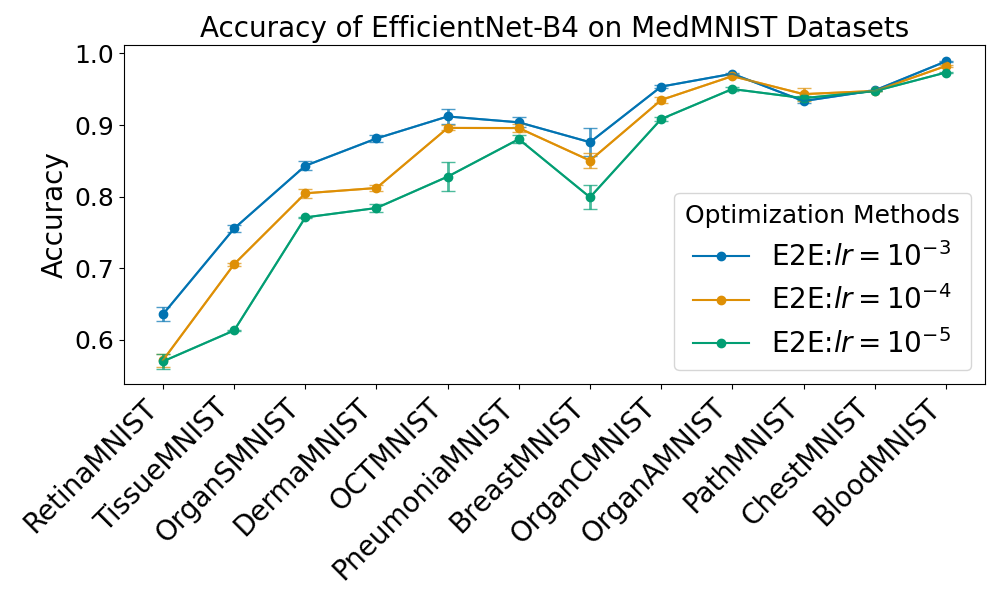}%
}
\end{figure}

\begin{figure}[h]
 % Caption and label go in the first argument and the figure contents
 % go in the second argument
\floatconts
  {fig:lr2}
  {\caption{Tuning the learning rate for the encoder of ViT-B/16 ranging in $\{10^{-4},10^{-5},10^{-6}\}$ on different datasets.}}
  {\includegraphics[width=0.7\linewidth]{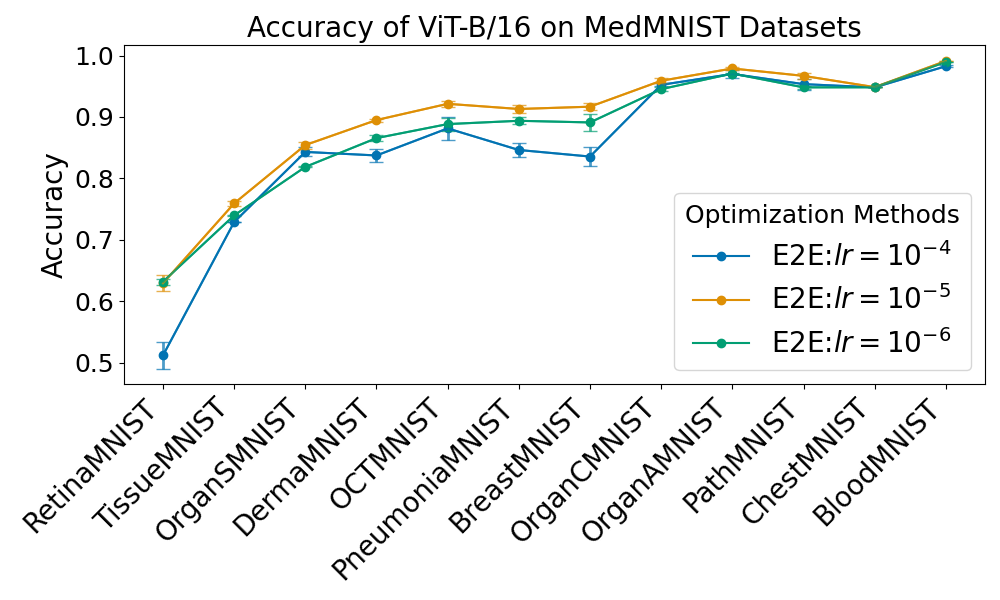}%
}
\end{figure}

\begin{figure}[h]
 % Caption and label go in the first argument and the figure contents
 % go in the second argument
\floatconts
  {fig:lr6}
  {\caption{Comparing the Accuracy of CLIP ViT-B/16 with the learning rate of the encoder ranging in $\{10^{-4},10^{-5},10^{-6}\}$ on different datasets.}}
  {
\includegraphics[width=0.7\linewidth]{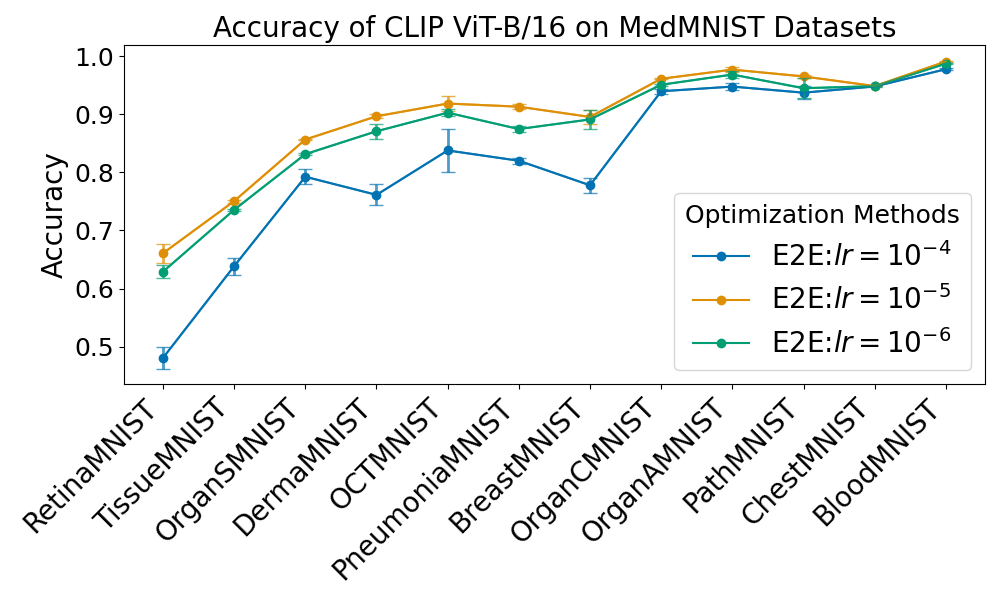}%
}
\end{figure}

\begin{figure}[h]
 % Caption and label go in the first argument and the figure contents
 % go in the second argument
\floatconts
  {fig:lr7}
  {\caption{Comparing the Accuracy of RSAM-C ViT-B/16 with the learning rate of the encoder ranging in $\{10^{-4},10^{-5},10^{-6}\}$ on different datasets.}}
  {
\includegraphics[width=0.7\linewidth]{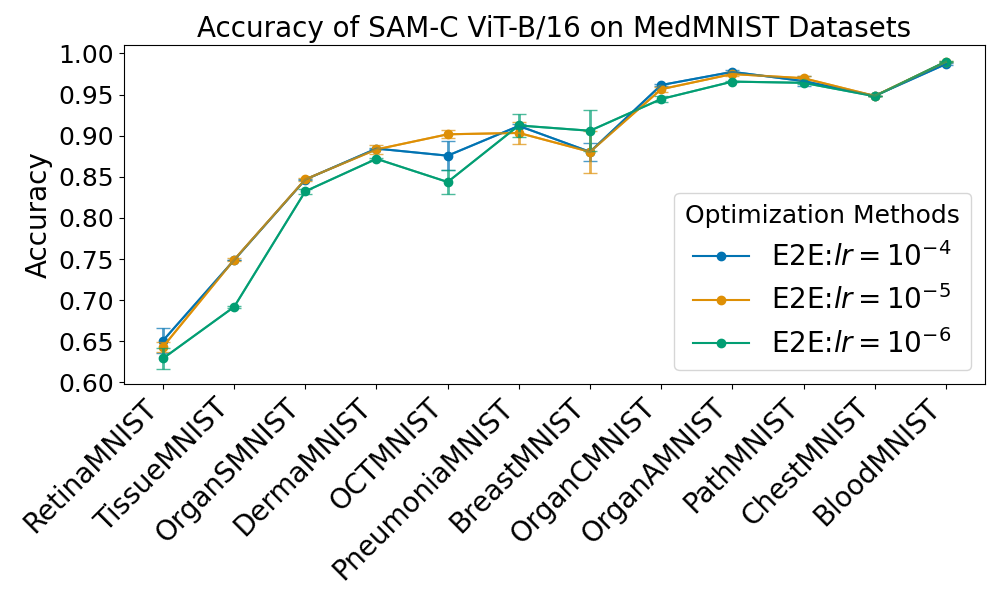}%
}
\end{figure}

\begin{figure}[h]
 % Caption and label go in the first argument and the figure contents
 % go in the second argument
\floatconts
  {fig:lr8}
  {\caption{Comparing the Accuracy of SAM ViT-B/16 with the learning rate of the encoder ranging in $\{10^{-4},10^{-5},10^{-6}\}$ on different datasets.}}
  {
\includegraphics[width=0.7\linewidth]{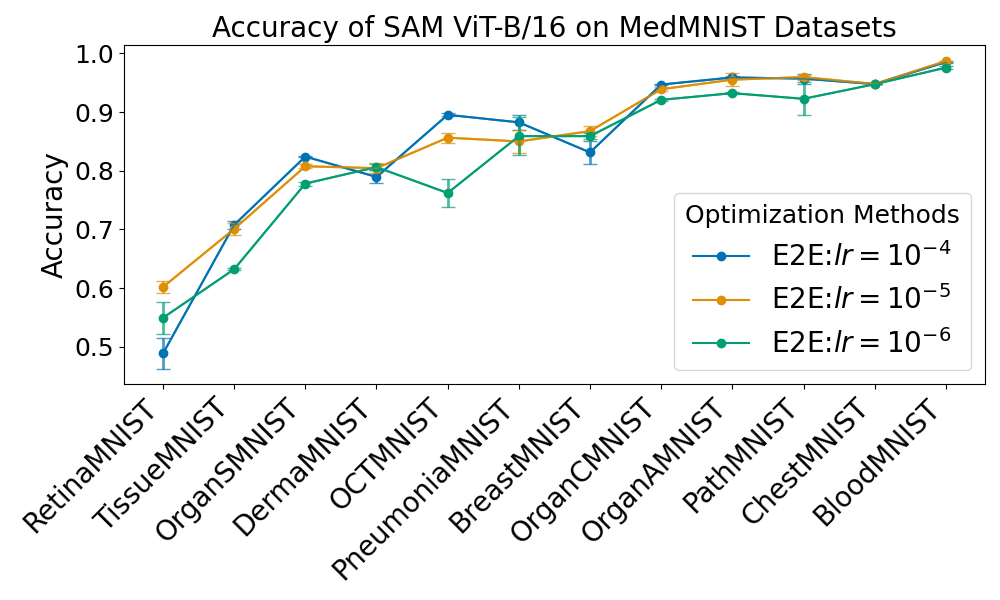}%
}
\end{figure}

\begin{figure}[h]
 % Caption and label go in the first argument and the figure contents
 % go in the second argument
\floatconts
  {fig:lr9}
  {\caption{Comparing the Accuracy of EVA-02 ViT-B/14 with the learning rate of the encoder ranging in $\{10^{-4},10^{-5},10^{-6}\}$ on different datasets.}}
  {
\includegraphics[width=0.7\linewidth]{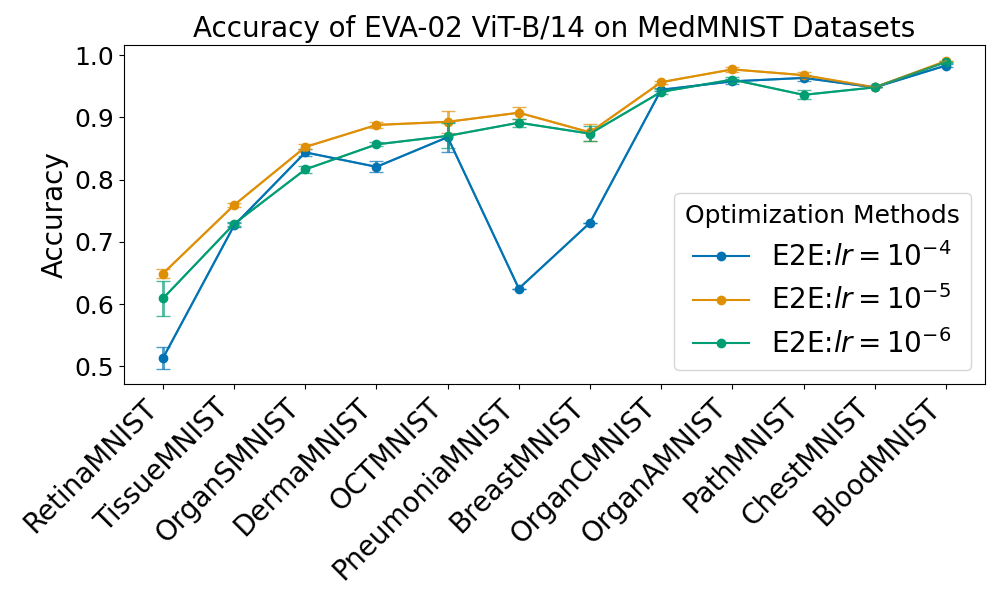}%
}
\end{figure}

\begin{figure}[h]
 % Caption and label go in the first argument and the figure contents
 % go in the second argument
\floatconts
  {fig:lr10}
  {\caption{Comparing the Accuracy of OpenAI ViT-B/16 with the learning rate of the encoder ranging in $\{10^{-4},10^{-5},10^{-6}\}$ on different datasets.}}
  {
\includegraphics[width=0.7\linewidth]{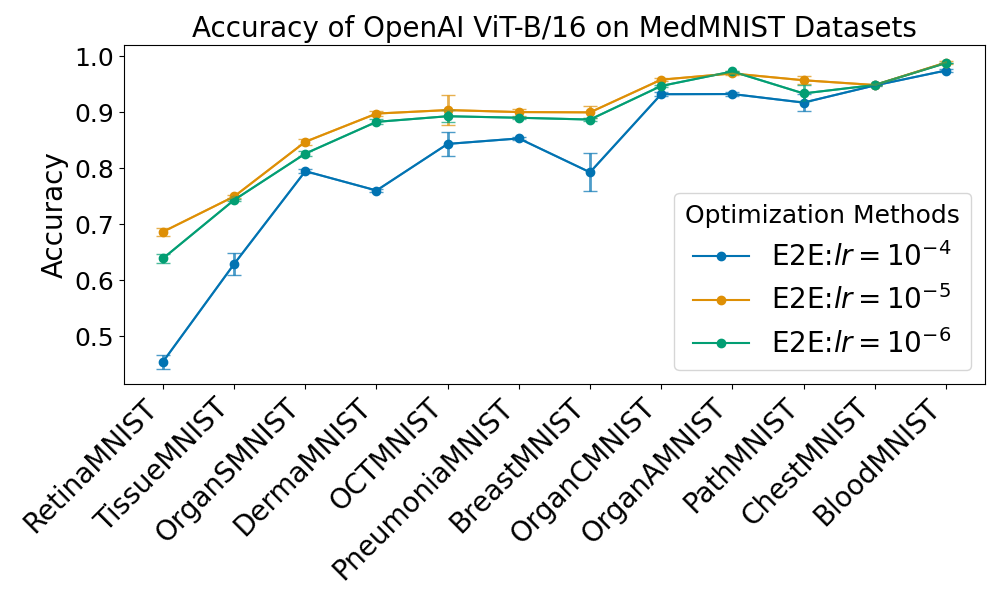}%
}
\end{figure}

\begin{figure}[h]
 % Caption and label go in the first argument and the figure contents
 % go in the second argument
\floatconts
  {fig:lr11}
  {\caption{Comparing the Accuracy of DINO ViT-B/16 with the learning rate of the encoder ranging in $\{10^{-4},10^{-5},10^{-6}\}$ on different datasets.}}
  {
\includegraphics[width=0.7\linewidth]{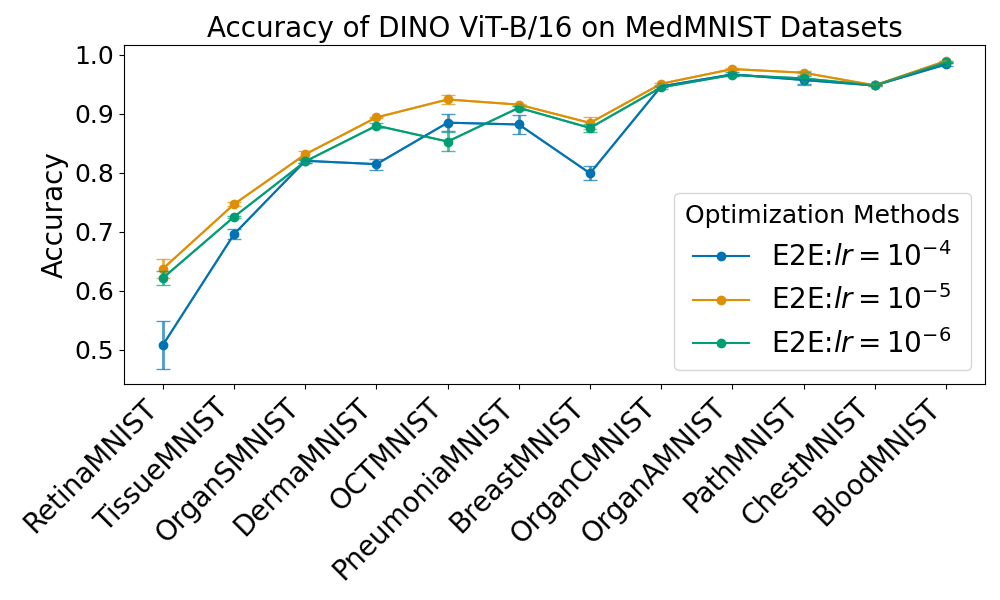}%
}
\end{figure}

\begin{figure}[h]
 % Caption and label go in the first argument and the figure contents
 % go in the second argument
\floatconts
  {fig:lr12}
  {\caption{Comparing the Accuracy of DINOv2 ViT-B/14 with the learning rate of the encoder ranging in $\{10^{-4},10^{-5},10^{-6}\}$ on different datasets.}}
  {
\includegraphics[width=0.7\linewidth]{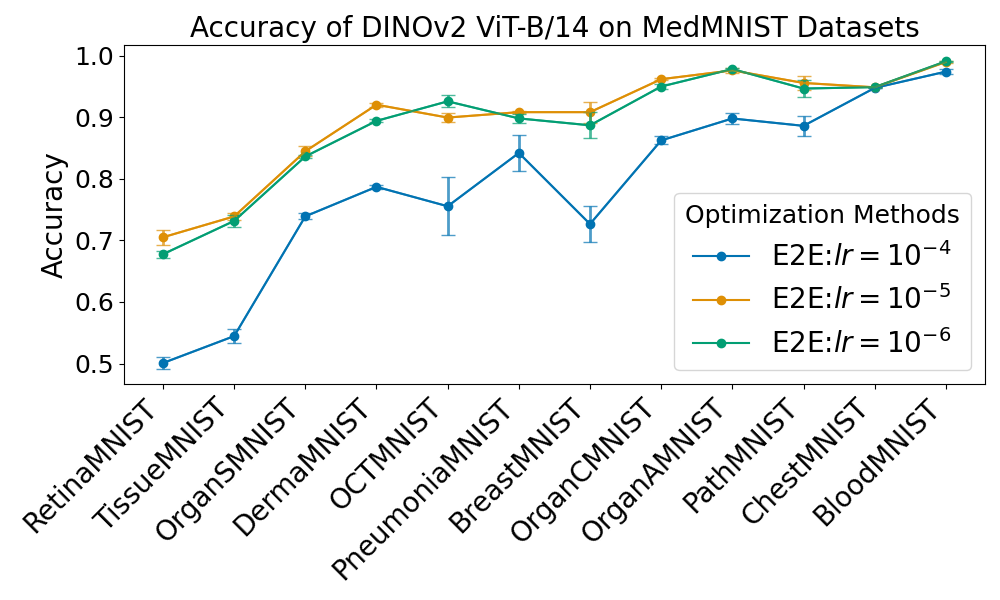}%
}
\end{figure}

\clearpage
\section{Results of Foundation Models on MedMNIST}\label{appendix:results_all}

\begin{savenotes}
\begin{table}[h]
\floatconts
  {tab:results2}%
  {\caption{Results of foundation models on BloodMNIST, ChestMNIST, and OCTMNIST.}}%
  {\scalebox{0.9}{
  \begin{tabular}{lcccccc}
  \hline
   & \multicolumn{2}{c}{BloodMNIST}& \multicolumn{2}{c}{ChestMNIST}& \multicolumn{2}{c}{OCTMNIST}\\
   \cmidrule(r){2-3}
   \cmidrule(r){4-5}
   \cmidrule(r){6-7}
   % \cmidrule(r){8-9}

   Methods &  ACC&  AUC&  ACC&  AUC& ACC &AUC \\
   \hline
   \multicolumn{7}{c}{END-TO-END}\\
   \hline
    VGG16& 98.90\tiny{$\pm$0.04} & 99.94\tiny{$\pm$0.01}  & 94.84\tiny{$\pm$0.02}  & 80.18\tiny{$\pm$0.97}  & 88.67\tiny{$\pm$0.34}  & 98.69\tiny{$\pm$0.22} \\
    DenseNet-121   & 99.06\tiny{$\pm$0.13} & 99.93\tiny{$\pm$0.00}  & 94.81\tiny{$\pm$0.01}  &  70.75\tiny{$\pm$0.84} & 90.63\tiny{$\pm$1.76}  &99.27\tiny{$\pm$0.30}  \\
    ResNet-18& 98.81\tiny{$\pm$0.23} & 99.93\tiny{$\pm$0.00}  & 94.82\tiny{$\pm$0.03}  & 80.13\tiny{$\pm$0.98}  &88.57\tiny{$\pm$0.90}   &99.01\tiny{$\pm$0.08}  \\
     EfficientNet-B4 &98.89\tiny{$\pm$0.10}  & \textbf{99.95\tiny{$\pm$0.00}}  & 94.84\tiny{$\pm$0.00}  & 82.65\tiny{$\pm$0.07}  &91.20\tiny{$\pm$0.99}   & 99.36\tiny{$\pm$0.28} \\
     ViT-B/16   & \textbf{99.11\tiny{$\pm$0.04}} & 99.95\tiny{$\pm$0.01}  & 94.85\tiny{$\pm$0.01}  &  81.28\tiny{$\pm$0.91} & 92.13\tiny{$\pm$0.46}  &99.63\tiny{$\pm$0.09}  \\
     CLIP ViT-B/16 &99.06\tiny{$\pm$0.13}  & 99.95\tiny{$\pm$0.01}  & 94.84\tiny{$\pm$0.00}  & 81.67\tiny{$\pm$0.91}  & 91.83\tiny{$\pm$1.36}  &99.55\tiny{$\pm$0.10}  \\
     SAM-C ViT-B/16  &99.01\tiny{$\pm$0.10}  & 99.91\tiny{$\pm$0.03}  & 94.84\tiny{$\pm$0.01}  & 71.17\tiny{$\pm$1.47}  & 90.17\tiny{$\pm$0.46}  & 99.22\tiny{$\pm$0.14} \\
     SAM ViT-B/16  &98.69\tiny{$\pm$0.06}  & 99.70\tiny{$\pm$0.07}  & 94.79\tiny{$\pm$0.00}  & 62.25\tiny{$\pm$1.00}  &89.53\tiny{$\pm$0.26}   &99.10\tiny{$\pm$0.38}  \\
    EVA-02 ViT-B/14   & 99.05\tiny{$\pm$0.06} & 99.68\tiny{$\pm$0.10}  & 94.83\tiny{$\pm$0.01}  & 67.06\tiny{$\pm$0.90}  &89.30\tiny{$\pm$1.73}   & 98.32\tiny{$\pm$0.82} \\
      OpenAI ViT-B/16   & 98.83\tiny{$\pm$0.33} & 99.95\tiny{$\pm$0.01}  & 94.84\tiny{$\pm$0.03}  & 76.18\tiny{$\pm$1.67}  & 90.37\tiny{$\pm$2.64}  & 99.54\tiny{$\pm$0.29} \\
    DINO ViT-B/16    & 98.97\tiny{$\pm$0.04} & 99.95\tiny{$\pm$0.01}  & 94.83\tiny{$\pm$0.03}  & 80.83\tiny{$\pm$0.28}  & 92.43\tiny{$\pm$0.79}  &99.53\tiny{$\pm$0.13}  \\
    % DINO ViT-S/16   & 99.03\tiny{$\pm$0.12} & 99.95\tiny{$\pm$0.01}  & 94.83\tiny{$\pm$0.03}  &  81.61\tiny{$\pm$0.10} & 85.70\tiny{$\pm$0.86}  & 99.09\tiny{$\pm$0.42} \\
    DINOv2 ViT-B/14  & 99.06\tiny{$\pm$0.10} & 99.94\tiny{$\pm$0.00}  &  \textbf{94.87\tiny{$\pm$0.01}} & \textbf{83.22\tiny{$\pm$0.48}}  & \textbf{92.57\tiny{$\pm$0.99}}  & \textbf{99.80\tiny{$\pm$0.02}} \\
    % DINOv2 ViT-S/14   & 99.18\tiny{$\pm$0.09} & 99.94\tiny{$\pm$0.00}  & 94.86\tiny{$\pm$0.01}  &  82.01\tiny{$\pm$0.42} & 90.17\tiny{$\pm$1.41}  & 99.71\tiny{$\pm$0.03} \\
    % MedSAM ViT-B/16     &  &   &   &   &   &  \\
    %  BiomedGPT  & \tiny{$\pm$}& \tiny{$\pm$} & \tiny{$\pm$} & \tiny{$\pm$} & \tiny{$\pm$} & \tiny{$\pm$}\\
           \hline
   \multicolumn{7}{c}{LINEAR PROBING}\\
   \hline
    VGG16& 91.87\tiny{$\pm$0.08} & 99.42\tiny{$\pm$0.00}  & 94.75\tiny{$\pm$0.00}  & 73.63\tiny{$\pm$0.14}  & 68.63\tiny{$\pm$0.25}  &96.20\tiny{$\pm$0.04}  \\
    DenseNet-121   & 95.95\tiny{$\pm$0.06} & 99.82\tiny{$\pm$0.00}  & 94.75\tiny{$\pm$0.00}  & 72.39\tiny{$\pm$0.87}  & 62.83\tiny{$\pm$0.21}  & 95.00\tiny{$\pm$0.06} \\
    ResNet-18& 92.84\tiny{$\pm$0.05} & 99.46\tiny{$\pm$0.01}  & 94.74\tiny{$\pm$0.00}  &  68.57\tiny{$\pm$0.01} &55.43\tiny{$\pm$0.17}   &91.52\tiny{$\pm$0.13}  \\
     EfficientNet-B4 & 91.71\tiny{$\pm$0.33} & 99.35\tiny{$\pm$0.01}  &94.70\tiny{$\pm$0.01}   & 66.39\tiny{$\pm$0.27}  & 63.27\tiny{$\pm$1.24}  &91.95\tiny{$\pm$0.33}  \\
     ViT-B/16   &98.12\tiny{$\pm$0.06}  & 99.93\tiny{$\pm$0.00}  & 94.73\tiny{$\pm$0.00}  & 74.61\tiny{$\pm$0.61}  &81.40\tiny{$\pm$0.28}   &98.82\tiny{$\pm$0.01}  \\
     CLIP ViT-B/16 & 96.45\tiny{$\pm$0.08} & 99.88\tiny{$\pm$0.00}  &94.75\tiny{$\pm$0.00}   &75.67\tiny{$\pm$0.06}   &79.67\tiny{$\pm$0.61}   &98.64\tiny{$\pm$0.01}  \\
     SAM-C ViT-B/16  & 97.14\tiny{$\pm$0.19} & 99.89\tiny{$\pm$0.01}  & \textbf{94.76\tiny{$\pm$0.01}}  & 75.86\tiny{$\pm$0.03}  & 75.47\tiny{$\pm$0.17}  &98.87\tiny{$\pm$0.01}  \\
     SAM ViT-B/16  &56.94\tiny{$\pm$0.10}  & 91.17\tiny{$\pm$0.02}  &94.74\tiny{$\pm$0.00}   &51.04\tiny{$\pm$0.11}   & 34.63\tiny{$\pm$0.05}  &74.18\tiny{$\pm$0.10}  \\
    EVA-02 ViT-B/14   & 96.49\tiny{$\pm$0.05} & 99.88\tiny{$\pm$0.00}  & 94.74\tiny{$\pm$0.00}  & 72.19\tiny{$\pm$0.08}  & 63.53\tiny{$\pm$0.34}  &97.13\tiny{$\pm$0.01}  \\
      OpenAI ViT-B/16   &95.50\tiny{$\pm$0.09}  & 99.79\tiny{$\pm$0.00}  & 94.74\tiny{$\pm$0.00}  &70.85\tiny{$\pm$0.49}   & 72.10\tiny{$\pm$0.65}  &96.44\tiny{$\pm$0.03}  \\
    DINO ViT-B/16    &\textbf{98.58\tiny{$\pm$0.04}}  &\textbf{99.94\tiny{$\pm$0.00}}   & \textbf{94.76\tiny{$\pm$0.01}}  &\textbf{78.82\tiny{$\pm$0.05}}   & 73.80\tiny{$\pm$0.43}  & 98.45\tiny{$\pm$0.06} \\
    % DINO ViT-S/16   & 98.28\tiny{$\pm$0.11} &99.95\tiny{$\pm$0.00}   &94.76\tiny{$\pm$0.01}   & 77.76\tiny{$\pm$0.28}  & 78.20\tiny{$\pm$0.28}  & 98.64\tiny{$\pm$0.09} \\
    DINOv2 ViT-B/14  & 97.80\tiny{$\pm$0.13} & 99.93\tiny{$\pm$0.00}  &94.74\tiny{$\pm$0.00}   &76.73\tiny{$\pm$0.04}   & \textbf{83.20\tiny{$\pm$0.16}}  & \textbf{99.11\tiny{$\pm$0.02}} \\
    % DINOv2 ViT-S/14   & 97.54\tiny{$\pm$0.13} & 99.92\tiny{$\pm$0.00}  & 94.73\tiny{$\pm$0.00}  &75.98\tiny{$\pm$0.10}   &80.50\tiny{$\pm$0.51}   & 99.07\tiny{$\pm$0.03} \\
    % MedSAM ViT-B/16     &  &   &   &   &   &  \\
    %  BiomedGPT  & \tiny{$\pm$}& \tiny{$\pm$} & \tiny{$\pm$} & \tiny{$\pm$} & \tiny{$\pm$} & \tiny{$\pm$}\\
           \hline
                         \multicolumn{7}{c}{Results from \citet{zhang2024generalist} for models pre-trained on medical data}\\
   \hline
BiomedGPT    & 98.7 &-  & 94.9 &-  &92.9  &-  \\
BiomedCLIP    & 97.9 &-  &93.0  &-  &85.6  &-  \\
MedSAM    & 97.2 & - & 83.8 &-  &89.1  & - \\

   \hline
  \end{tabular}}
  }
\end{table}
\end{savenotes}

\begin{savenotes}
\begin{table}[h]
\floatconts
  {tab:results3}%
  {\caption{Results of foundation models on OrganAMNIST, OrganCMNIST, and OrganSMNIST.}}%
  {\scalebox{0.9}{
  \begin{tabular}{lcccccc}
  \hline
   & \multicolumn{2}{c}{OrganAMNIST}& \multicolumn{2}{c}{OrganCMNIST}& \multicolumn{2}{c}{OrganSMNIST}\\
   \cmidrule(r){2-3}
   \cmidrule(r){4-5}
   \cmidrule(r){6-7}
   % \cmidrule(r){8-9}

   Methods &  ACC&  AUC&  ACC&  AUC& ACC &AUC \\
   \hline
   \multicolumn{7}{c}{END-TO-END}\\
   \hline
    VGG16& 96.02\tiny{$\pm$0.44} & 99.87\tiny{$\pm$0.03}  &93.41\tiny{$\pm$1.38}   &99.54\tiny{$\pm$0.20}   & 81.86\tiny{$\pm$0.74}  &97.24\tiny{$\pm$0.38}  \\
    DenseNet-121   & 97.52\tiny{$\pm$0.25} & 99.93\tiny{$\pm$0.02}  &95.42\tiny{$\pm$0.09}   & 99.80\tiny{$\pm$0.02}  & 85.28\tiny{$\pm$0.27}  & 98.05\tiny{$\pm$0.06} \\
    ResNet-18& 96.66\tiny{$\pm$0.12} &99.85\tiny{$\pm$0.06}   &94.90\tiny{$\pm$0.71}   &99.76\tiny{$\pm$0.04}   & 84.81\tiny{$\pm$0.28}  &97.52\tiny{$\pm$0.01}  \\
     EfficientNet-B4 & 97.17\tiny{$\pm$0.09} &99.83\tiny{$\pm$0.03}   & 95.37\tiny{$\pm$0.26}  & 99.79\tiny{$\pm$0.03}  &84.32\tiny{$\pm$0.65}   &98.17\tiny{$\pm$0.12}  \\
     ViT-B/16   &\textbf{97.89\tiny{$\pm$0.25}}  & \textbf{99.96\tiny{$\pm$0.01}}  &95.89\tiny{$\pm$0.43}   & 99.88\tiny{$\pm$0.01}  &85.42\tiny{$\pm$0.58}   & 98.33\tiny{$\pm$0.05} \\
     CLIP ViT-B/16 &97.67\tiny{$\pm$0.37}  &99.93\tiny{$\pm$0.01}   & 96.09\tiny{$\pm$0.08}  &\textbf{99.89\tiny{$\pm$0.00} }  &\textbf{85.65\tiny{$\pm$0.13}}   & 98.26\tiny{$\pm$0.07} \\
     SAM-C ViT-B/16  &97.76\tiny{$\pm$0.17}  & 99.91\tiny{$\pm$0.01}  & 96.15\tiny{$\pm$0.08}  &99.72\tiny{$\pm$0.07}   & 84.69\tiny{$\pm$0.20}  & 96.96\tiny{$\pm$0.40} \\
     SAM ViT-B/16  & 95.90\tiny{$\pm$0.29} &99.63\tiny{$\pm$0.15}   & 94.69\tiny{$\pm$0.09}  &99.01\tiny{$\pm$0.12}   & 82.41\tiny{$\pm$0.10}  & 97.01\tiny{$\pm$0.27} \\
    EVA-02 ViT-B/14   &97.72\tiny{$\pm$0.35}  & 99.34\tiny{$\pm$0.16}  &  95.67\tiny{$\pm$0.25} & 98.37\tiny{$\pm$0.09}  & 85.25\tiny{$\pm$0.40}  &91.42\tiny{$\pm$1.46}  \\
      OpenAI ViT-B/16   &97.25\tiny{$\pm$0.04}  &99.90\tiny{$\pm$0.01}   & 95.78\tiny{$\pm$0.29}  &99.81\tiny{$\pm$0.01}   & 84.71\tiny{$\pm$0.53}  & 98.01\tiny{$\pm$0.18} \\
    DINO ViT-B/16    & 97.59\tiny{$\pm$0.16} &99.91\tiny{$\pm$0.02}   & 95.10\tiny{$\pm$0.14}  & 99.83\tiny{$\pm$0.01}  & 83.15\tiny{$\pm$0.56}  & 98.05\tiny{$\pm$0.16} \\
    % DINO ViT-S/16   &97.46\tiny{$\pm$0.15}  & 99.93\tiny{$\pm$0.02}  &95.11\tiny{$\pm$0.28}   & 99.82\tiny{$\pm$0.03}  &  84.43\tiny{$\pm$0.28} & 98.22\tiny{$\pm$0.02} \\
    DINOv2 ViT-B/14  & 97.81\tiny{$\pm$0.09} &99.91\tiny{$\pm$0.01}   & \textbf{96.16\tiny{$\pm$0.20}}  & 99.88\tiny{$\pm$0.01}  & 84.46\tiny{$\pm$0.82}  &\textbf{98.41\tiny{$\pm$0.07}}  \\
    % DINOv2 ViT-S/14   & 97.60\tiny{$\pm$0.07} &99.93\tiny{$\pm$0.02}   &  95.92\tiny{$\pm$0.18} & 99.85\tiny{$\pm$0.03}  & 84.40\tiny{$\pm$0.71}  & 98.16\tiny{$\pm$0.13} \\
    % MedSAM ViT-B/16     &  &   &   &   &   &  \\
    %  BiomedGPT  & \tiny{$\pm$}& \tiny{$\pm$} & \tiny{$\pm$} & \tiny{$\pm$} & \tiny{$\pm$} & \tiny{$\pm$}\\
           \hline
   \multicolumn{7}{c}{LINEAR PROBING}\\
   \hline
    VGG16&91.56\tiny{$\pm$0.07}  & 99.54\tiny{$\pm$0.00}  &85.83\tiny{$\pm$0.09}   & 98.85\tiny{$\pm$0.00}  & 73.07\tiny{$\pm$0.08}  & 96.58\tiny{$\pm$0.01} \\
    DenseNet-121   &91.53\tiny{$\pm$0.11}  & 99.52\tiny{$\pm$0.01}  & 86.45\tiny{$\pm$0.02}  & 99.05\tiny{$\pm$0.01}  & 74.86\tiny{$\pm$0.20}  & 97.01\tiny{$\pm$0.02} \\
    ResNet-18& 87.05\tiny{$\pm$0.20} &98.95\tiny{$\pm$0.01}   & 82.57\tiny{$\pm$0.22}  & 98.29\tiny{$\pm$0.02}  & 69.79\tiny{$\pm$0.16}  & 95.95\tiny{$\pm$0.02} \\
     EfficientNet-B4 &87.51\tiny{$\pm$0.09}  &98.99\tiny{$\pm$0.01}   & 82.26\tiny{$\pm$0.06}  &98.46\tiny{$\pm$0.01}   &69.32\tiny{$\pm$0.16}   & 95.77\tiny{$\pm$0.03} \\
     ViT-B/16   & 93.02\tiny{$\pm$0.03} & 99.65\tiny{$\pm$0.00}  &88.42\tiny{$\pm$0.05}   &99.22\tiny{$\pm$0.01}   & 77.66\tiny{$\pm$0.12}  & 97.59\tiny{$\pm$0.01} \\
     CLIP ViT-B/16 &90.87\tiny{$\pm$0.09}  & 99.44\tiny{$\pm$0.01}  & 85.14\tiny{$\pm$0.12}  &98.84\tiny{$\pm$0.00}   & 74.94\tiny{$\pm$0.26}  &96.94\tiny{$\pm$0.03}  \\
     SAM-C ViT-B/16  & 93.58\tiny{$\pm$0.05} & 99.68\tiny{$\pm$0.00}  & 89.67\tiny{$\pm$0.07}  & 99.37\tiny{$\pm$0.00}  & 78.93\tiny{$\pm$0.07}  & 97.84\tiny{$\pm$0.00} \\
     SAM ViT-B/16  & 68.69\tiny{$\pm$0.07} &94.76\tiny{$\pm$0.01}   & 68.29\tiny{$\pm$0.07}  &  94.25\tiny{$\pm$0.01} &54.81\tiny{$\pm$0.01}   &90.59\tiny{$\pm$0.01}  \\
    EVA-02 ViT-B/14   &91.51\tiny{$\pm$0.03}  & 99.55\tiny{$\pm$0.00}  & 86.90\tiny{$\pm$0.26}  & 99.00\tiny{$\pm$0.01}  & 77.32\tiny{$\pm$0.16}  & 97.49\tiny{$\pm$0.03} \\
      OpenAI ViT-B/16   & 88.40\tiny{$\pm$0.07} & 99.17\tiny{$\pm$0.00}  & 82.66\tiny{$\pm$0.03}  & 98.48\tiny{$\pm$0.00}  &74.19\tiny{$\pm$0.38}   &96.82\tiny{$\pm$0.06}  \\
    DINO ViT-B/16    &\textbf{94.68\tiny{$\pm$0.06}}  & \textbf{99.79\tiny{$\pm$0.01}}  & \textbf{91.63\tiny{$\pm$0.21}}  &  \textbf{99.56\tiny{$\pm$0.00}} & \textbf{79.42\tiny{$\pm$0.07}}  & \textbf{97.85\tiny{$\pm$0.00}} \\
    % DINO ViT-S/16   & 94.21\tiny{$\pm$0.10} & 99.68\tiny{$\pm$0.00}  & 90.03\tiny{$\pm$0.07}  & 99.44\tiny{$\pm$0.01}  & 78.51\tiny{$\pm$0.15}  & 97.80\tiny{$\pm$0.00} \\
    DINOv2 ViT-B/14  & 91.99\tiny{$\pm$0.11} & 99.54\tiny{$\pm$0.01}  & 87.03\tiny{$\pm$0.15}  & 98.96\tiny{$\pm$0.01}  & 76.66\tiny{$\pm$0.16}  & 97.38\tiny{$\pm$0.01} \\
    % DINOv2 ViT-S/14   &92.91\tiny{$\pm$0.09}  & 99.66\tiny{$\pm$0.00}  &88.17\tiny{$\pm$0.13}   & 99.10\tiny{$\pm$0.02}  & 77.47\tiny{$\pm$0.32}  & 97.53\tiny{$\pm$0.01} \\
    % MedSAM ViT-B/16     &  &   &   &   &   &  \\
    %  BiomedGPT  & \tiny{$\pm$}& \tiny{$\pm$} & \tiny{$\pm$} & \tiny{$\pm$} & \tiny{$\pm$} & \tiny{$\pm$}\\
           \hline
                                    \multicolumn{7}{c}{Results from \citet{zhang2024generalist} for models pre-trained on medical data}\\
   \hline
BiomedGPT    & - &-  &91.0  &-  &-  &-  \\
BiomedCLIP    &-  &-  & 92.5 &-&-& - \\
MedSAM    &-  & - & 90.6 &-  &-  &-  \\

   \hline
  \end{tabular}}
  }
\end{table}
\end{savenotes}

\begin{savenotes}
\begin{table}[h]
\floatconts
  {tab:results4}%
  {\caption{Results of foundation models on PathMNIST, RetinaMNIST, and TissueMNIST.}}%
  {\scalebox{0.9}{
  \begin{tabular}{lcccccc}
  \hline
   & \multicolumn{2}{c}{PathMNIST}& \multicolumn{2}{c}{RetinaMNIST}& \multicolumn{2}{c}{TissueMNIST}\\
   \cmidrule(r){2-3}
   \cmidrule(r){4-5}
   \cmidrule(r){6-7}
   % \cmidrule(r){8-9}

   Methods &  ACC&  AUC&  ACC&  AUC& ACC &AUC \\
   \hline
   \multicolumn{7}{c}{END-TO-END}\\
   \hline
    VGG16& 95.22\tiny{$\pm$0.55} & 99.58\tiny{$\pm$0.13}  &66.00\tiny{$\pm$1.59}   & 85.17\tiny{$\pm$1.49}  &71.08\tiny{$\pm$0.46}   & 94.17\tiny{$\pm$0.05} \\
    DenseNet-121   &95.06\tiny{$\pm$1.00}  & 99.72\tiny{$\pm$0.03}  &65.17\tiny{$\pm$2.82}   &82.10\tiny{$\pm$0.84}   & 73.84\tiny{$\pm$0.34}  & 95.22\tiny{$\pm$0.07} \\
    ResNet-18& 94.40\tiny{$\pm$1.41} &99.64\tiny{$\pm$0.16}   & 62.33\tiny{$\pm$1.05}  & 80.55\tiny{$\pm$0.34}  &71.70\tiny{$\pm$0.37}   & 94.44\tiny{$\pm$0.06} \\
     EfficientNet-B4 &94.32\tiny{$\pm$0.90}  &99.61\tiny{$\pm$0.08}   &63.58\tiny{$\pm$0.96}   & 83.42\tiny{$\pm$0.69}  &75.56\tiny{$\pm$0.44}   & 95.69\tiny{$\pm$0.10} \\
     ViT-B/16   &96.69\tiny{$\pm$0.52}  & 99.72\tiny{$\pm$0.02}  &63.17\tiny{$\pm$0.51}   & 83.11\tiny{$\pm$0.77}  & 75.92\tiny{$\pm$0.38}  & \textbf{95.85\tiny{$\pm$0.11} }\\
     CLIP ViT-B/16 &96.50\tiny{$\pm$0.24}  & 99.75\tiny{$\pm$0.01}  & 66.08\tiny{$\pm$1.64}  & 85.79\tiny{$\pm$1.18}  &  75.05\tiny{$\pm$0.27} &95.47\tiny{$\pm$0.10}  \\
     SAM-C ViT-B/16  & \textbf{96.99\tiny{$\pm$0.25}} & 99.45\tiny{$\pm$0.11}  & 65.08\tiny{$\pm$1.48}  & 80.36\tiny{$\pm$2.33}  & 74.92\tiny{$\pm$0.15}  &95.41\tiny{$\pm$0.08}  \\
     SAM ViT-B/16  &95.97\tiny{$\pm$0.43}  & 98.66\tiny{$\pm$0.66}  & 60.17\tiny{$\pm$1.05}  & 76.07\tiny{$\pm$1.93}  & 70.76\tiny{$\pm$0.70}  & 94.10\tiny{$\pm$0.30} \\
    EVA-02 ViT-B/14   & 96.80\tiny{$\pm$0.43} & 98.64\tiny{$\pm$0.25}  & 64.92\tiny{$\pm$0.66}  & 78.92\tiny{$\pm$2.48}  & \textbf{75.92\tiny{$\pm$0.28}}  &92.91\tiny{$\pm$0.24}  \\
      OpenAI ViT-B/16   & 95.68\tiny{$\pm$0.72} & 99.65\tiny{$\pm$0.10}  &68.67\tiny{$\pm$0.72}   & 84.91\tiny{$\pm$1.33}  &74.96\tiny{$\pm$0.21}   &95.52\tiny{$\pm$0.09}  \\
    DINO ViT-B/16    & 96.97\tiny{$\pm$0.33} & \textbf{99.75\tiny{$\pm$0.01}}  & 63.75\tiny{$\pm$1.62}  &82.60\tiny{$\pm$0.79}   & 74.69\tiny{$\pm$0.34}  & 95.43\tiny{$\pm$0.11} \\
    % DINO ViT-S/16   &96.61\tiny{$\pm$0.46}  & 99.71\tiny{$\pm$0.04}  & 62.25\tiny{$\pm$1.95}  &83.04\tiny{$\pm$1.70}   & 74.13\tiny{$\pm$0.25}  & 95.26\tiny{$\pm$0.03} \\
    DINOv2 ViT-B/14  &95.55\tiny{$\pm$1.14}  &99.59\tiny{$\pm$0.19}   & \textbf{70.50\tiny{$\pm$1.24}}  & \textbf{89.26\tiny{$\pm$0.63}}  & 73.88\tiny{$\pm$0.61}  & 95.29\tiny{$\pm$0.13} \\
    % DINOv2 ViT-S/14   & 94.90\tiny{$\pm$0.08} & 99.53\tiny{$\pm$0.06}  & 69.75\tiny{$\pm$2.35}  & 86.60\tiny{$\pm$1.47}  & 74.94\tiny{$\pm$0.32}  & 95.56\tiny{$\pm$0.10} \\
    % MedSAM ViT-B/16     &  &   &   &   &   &  \\
    %  BiomedGPT  & \tiny{$\pm$}& \tiny{$\pm$} & \tiny{$\pm$} & \tiny{$\pm$} & \tiny{$\pm$} & \tiny{$\pm$}\\
           \hline
   \multicolumn{7}{c}{LINEAR PROBING}\\
   \hline
    VGG16&88.23\tiny{$\pm$0.08}  & 98.81\tiny{$\pm$0.01}   & 59.83\tiny{$\pm$1.53}  & 83.96\tiny{$\pm$0.49}  &57.21\tiny{$\pm$0.04}   & 87.30\tiny{$\pm$0.01} \\
    DenseNet-121   & 93.30\tiny{$\pm$0.22} &99.35\tiny{$\pm$0.03}   & 62.83\tiny{$\pm$0.62}  & 83.34\tiny{$\pm$0.44}  &  55.96\tiny{$\pm$0.04} & 86.25\tiny{$\pm$0.01} \\
    ResNet-18&  90.58\tiny{$\pm$0.34}& 99.22\tiny{$\pm$0.01}  &61.08\tiny{$\pm$0.62}   & 82.26\tiny{$\pm$0.16}  &52.20\tiny{$\pm$0.07}   & 83.99\tiny{$\pm$0.02} \\
     EfficientNet-B4 & 88.94\tiny{$\pm$0.25} & 98.77\tiny{$\pm$0.09}  &57.33\tiny{$\pm$1.39}   & 77.33\tiny{$\pm$1.09}  & 49.31\tiny{$\pm$0.08}  & 80.82\tiny{$\pm$0.09} \\
     ViT-B/16   & 93.68\tiny{$\pm$0.17} & 99.64\tiny{$\pm$0.01}  & 59.92\tiny{$\pm$0.77}  & 83.89\tiny{$\pm$0.18}  & 62.24\tiny{$\pm$0.01}  & 90.08\tiny{$\pm$0.02} \\
     CLIP ViT-B/16 & 91.30\tiny{$\pm$0.16} &99.24\tiny{$\pm$0.02}   &62.00\tiny{$\pm$0.61}   & 85.85\tiny{$\pm$0.10}  & 60.34\tiny{$\pm$0.09}  & 89.07\tiny{$\pm$0.01} \\
     SAM-C ViT-B/16  &93.74\tiny{$\pm$0.06}  &  99.51\tiny{$\pm$0.01} &62.92\tiny{$\pm$0.47}   & 85.69\tiny{$\pm$0.05}  & 59.32\tiny{$\pm$0.03}  &88.59\tiny{$\pm$0.01}  \\
     SAM ViT-B/16  & 66.06\tiny{$\pm$0.09} & 93.42\tiny{$\pm$0.01}  & 43.50\tiny{$\pm$0.00}  & 55.26\tiny{$\pm$3.93}  & 43.98\tiny{$\pm$0.00}  &  72.42\tiny{$\pm$0.02}\\
    EVA-02 ViT-B/14   & 93.53\tiny{$\pm$0.07} &99.60\tiny{$\pm$0.00}   & 66.00\tiny{$\pm$0.20}  & 86.11\tiny{$\pm$0.01}  &  59.72\tiny{$\pm$0.05} & 88.75\tiny{$\pm$0.01} \\
      OpenAI ViT-B/16   &93.27\tiny{$\pm$0.05}  & 99.52\tiny{$\pm$0.01}  &62.75\tiny{$\pm$0.54}   & 84.10\tiny{$\pm$0.04}  &  57.94\tiny{$\pm$0.04} & 87.62\tiny{$\pm$0.02} \\
    DINO ViT-B/16    & \textbf{96.20\tiny{$\pm$0.09}} &  \textbf{99.78\tiny{$\pm$0.01}} &63.25\tiny{$\pm$1.67}   &85.27\tiny{$\pm$0.10}   & \textbf{62.78\tiny{$\pm$0.08}}  & \textbf{90.33\tiny{$\pm$0.01}} \\
    % DINO ViT-S/16   & 94.66\tiny{$\pm$0.17}  &99.72\tiny{$\pm$0.00}   & 62.08\tiny{$\pm$0.62}  & 83.84\tiny{$\pm$0.11}  &  61.24\tiny{$\pm$0.02} & 89.49\tiny{$\pm$0.05} \\
    DINOv2 ViT-B/14  & 93.82\tiny{$\pm$0.07} & 99.49\tiny{$\pm$0.00}  & \textbf{69.08\tiny{$\pm$0.66}}  &\textbf{88.12\tiny{$\pm$0.47}}   & 61.98\tiny{$\pm$0.06}  & 89.92\tiny{$\pm$0.01} \\
    % DINOv2 ViT-S/14   & 93.28\tiny{$\pm$0.02} & 99.48\tiny{$\pm$0.01}  &65.58\tiny{$\pm$0.31}   &86.49\tiny{$\pm$0.21}   & 60.82\tiny{$\pm$0.04}  & 89.29\tiny{$\pm$0.01} \\
    % MedSAM ViT-B/16     &  &   &   &   &   &  \\
    %  BiomedGPT  & \tiny{$\pm$}& \tiny{$\pm$} & \tiny{$\pm$} & \tiny{$\pm$} & \tiny{$\pm$} & \tiny{$\pm$}\\
           \hline
                                    \multicolumn{7}{c}{Results from \citet{zhang2024generalist} for models pre-trained on medical data}\\
   \hline
BiomedGPT    &95.8  &-  &-  &-  &-  &-\\
BiomedCLIP    &91.0  &-  &-  &-  & - &-  \\
MedSAM    & 84.2 & - & - &  -&-& - \\

   \hline
  \end{tabular}}
  }
\end{table}
\end{savenotes}

\clearpage
\section{Results of Foundation Models on DermaMNIST with Different Image sizes}\label{appendix:results_derma_sizes}

\begin{savenotes}
\begin{table}[h]
\floatconts
  {tab:results_sizes_pad}%
  {\caption{Results of foundation models on DermaMNIST of different image sizes using zero-padding to resize the original images into a resolution of $224\times 224$.}}%
  {\scalebox{0.9}{
  \begin{tabular}{lcccccc}
  \hline
   & \multicolumn{2}{c}{$28\times 28$}& \multicolumn{2}{c}{$64\times 64$}& \multicolumn{2}{c}{$128\times 128$}\\
   \cmidrule(r){2-3}
   \cmidrule(r){4-5}
   \cmidrule(r){6-7}
   % \cmidrule(r){8-9}

   Methods &  ACC&  AUC&  ACC&  AUC& ACC &AUC \\
   \hline
   \multicolumn{7}{c}{END-TO-END}\\
   \hline
    VGG16 &79.62\tiny{$\pm$0.94}  &91.98\tiny{$\pm$0.46}   & \textbf{84.26\tiny{$\pm$0.58}}  & 94.93\tiny{$\pm$0.45}  & 86.80\tiny{$\pm$0.21}  & 96.96\tiny{$\pm$0.18} \\
    DenseNet-121    & 76.11\tiny{$\pm$0.57} &87.65\tiny{$\pm$1.24}   & 81.86\tiny{$\pm$0.36}  &93.76\tiny{$\pm$0.63}   & 85.60\tiny{$\pm$0.19}  & 95.69\tiny{$\pm$0.30} \\
    ResNet-18 &77.71\tiny{$\pm$0.71}  &90.87\tiny{$\pm$0.75}   & 81.16\tiny{$\pm$1.18}  &93.34\tiny{$\pm$1.11}   &84.22\tiny{$\pm$0.68}   & 95.26\tiny{$\pm$0.13} \\
     EfficientNet-B4  &73.85\tiny{$\pm$0.56}  &88.40\tiny{$\pm$0.75}   & 78.72\tiny{$\pm$0.82}  &92.51\tiny{$\pm$0.75}   & 84.74\tiny{$\pm$1.63}  & 95.87\tiny{$\pm$0.47} \\
     ViT-B/16    &78.77\tiny{$\pm$0.16}  & 92.82\tiny{$\pm$0.40}  & 83.52\tiny{$\pm$0.89}  &\textbf{96.13\tiny{$\pm$0.27}}   &86.32\tiny{$\pm$0.24}   &97.37\tiny{$\pm$0.07}  \\
     CLIP ViT-B/16  &\textbf{79.67\tiny{$\pm$0.80}}  &92.57\tiny{$\pm$0.22}   & 82.96\tiny{$\pm$0.69}  &95.45\tiny{$\pm$0.52}   & 86.93\tiny{$\pm$0.19}  &97.51\tiny{$\pm$0.07}  \\
     SAM-C ViT-B/16   & 77.37\tiny{$\pm$0.76} & 89.25\tiny{$\pm$0.75}  &81.00\tiny{$\pm$1.40}   & 93.48\tiny{$\pm$0.45}  & 85.07\tiny{$\pm$0.20}  & 95.30\tiny{$\pm$0.18} \\
     SAM ViT-B/16   &73.95\tiny{$\pm$0.48}  & 79.83\tiny{$\pm$0.90}  &75.79\tiny{$\pm$1.12}   &84.00\tiny{$\pm$2.90}   & 78.34\tiny{$\pm$0.36}  & 87.21\tiny{$\pm$1.31} \\
    EVA-02 ViT-B/14    & 78.87\tiny{$\pm$0.72} &83.27\tiny{$\pm$1.32}  & 82.11\tiny{$\pm$0.35}  & 87.51\tiny{$\pm$1.17}  &85.89\tiny{$\pm$0.25}   &92.17\tiny{$\pm$2.00}  \\
      OpenAI ViT-B/16    & 78.94\tiny{$\pm$0.49} &92.45\tiny{$\pm$0.13}   & 83.28\tiny{$\pm$0.26}  &95.44\tiny{$\pm$0.29}   &87.22\tiny{$\pm$0.59}   &97.12\tiny{$\pm$0.20}  \\
    DINO ViT-B/16     &78.35\tiny{$\pm$0.11}  &\textbf{93.14\tiny{$\pm$0.19}}   &81.83\tiny{$\pm$0.49}   & 95.49\tiny{$\pm$0.37}  &85.34\tiny{$\pm$0.43}   &97.10\tiny{$\pm$0.06}  \\
    % DINO ViT-S/16    &  &   &   &   &   &  \\
    DINOv2 ViT-B/14  &76.79\tiny{$\pm$0.72}  &92.13\tiny{$\pm$0.54}   & 81.68\tiny{$\pm$0.76}  &94.96\tiny{$\pm$0.67}   &\textbf{87.35\tiny{$\pm$0.06}}   &\textbf{97.60\tiny{$\pm$0.02}}  \\
    % DINOv2 ViT-S/14    &  &   &   &   &   &  \\
    % MedSAM ViT-B/16     &  &   &   &   &   &  \\
    %  BiomedGPT   &  &   &   &   &   &  \\
           \hline
   \multicolumn{7}{c}{LINEAR PROBING}\\
   \hline
    VGG16 & 73.47\tiny{$\pm$0.18} & 88.76\tiny{$\pm$0.07}  & 75.79\tiny{$\pm$0.27}  &89.70\tiny{$\pm$0.08}   &76.86\tiny{$\pm$0.07}   &91.29\tiny{$\pm$0.02}  \\
    DenseNet-121    &70.27\tiny{$\pm$0.07}  &85.13\tiny{$\pm$1.33}   &71.80\tiny{$\pm$0.16}   & 87.33\tiny{$\pm$0.03}  &75.98\tiny{$\pm$0.33}   &91.34\tiny{$\pm$0.14}  \\
    ResNet-18 & 69.68\tiny{$\pm$0.19} & 82.37\tiny{$\pm$0.19}  & 68.43\tiny{$\pm$0.15}  &83.23\tiny{$\pm$0.04}   &72.92\tiny{$\pm$0.24}   &88.34\tiny{$\pm$0.03}  \\
     EfficientNet-B4  & 68.23\tiny{$\pm$0.15} &81.28\tiny{$\pm$0.45}   &70.09\tiny{$\pm$0.20}   &84.62\tiny{$\pm$0.26}   &72.14\tiny{$\pm$0.13}   &86.97\tiny{$\pm$0.57}  \\
     ViT-B/16    & 72.52\tiny{$\pm$0.14} &90.74\tiny{$\pm$0.17}   & 78.59\tiny{$\pm$0.30}  &94.18\tiny{$\pm$0.04}   & 81.91\tiny{$\pm$0.22}  & 95.54\tiny{$\pm$0.03} \\
     CLIP ViT-B/16  &74.58\tiny{$\pm$0.24}  &90.47\tiny{$\pm$0.16}   &77.27\tiny{$\pm$0.19}   & 93.54\tiny{$\pm$0.05}  & 81.23\tiny{$\pm$0.22}  & 95.13\tiny{$\pm$0.02} \\
     SAM-C ViT-B/16   &73.67\tiny{$\pm$0.07}  &90.16\tiny{$\pm$0.06}   &78.99\tiny{$\pm$0.25}   &94.59\tiny{$\pm$0.02}   & 80.45\tiny{$\pm$0.08}  & 95.57\tiny{$\pm$0.03} \\
     SAM ViT-B/16   &66.88\tiny{$\pm$0.00}  &55.42\tiny{$\pm$2.35}   & 66.88\tiny{$\pm$0.00}  &46.89\tiny{$\pm$2.67}   & 66.88\tiny{$\pm$0.00}  &52.40\tiny{$\pm$1.05}  \\
    EVA-02 ViT-B/14    &75.48\tiny{$\pm$0.06}  & 92.40\tiny{$\pm$0.03}  & 78.89\tiny{$\pm$0.27}  &94.28\tiny{$\pm$0.06}   &81.68\tiny{$\pm$0.10}   &95.31\tiny{$\pm$0.01}  \\
      OpenAI ViT-B/16    & 75.81\tiny{$\pm$0.11} &91.82\tiny{$\pm$0.05}  &79.62\tiny{$\pm$0.10}   &94.36\tiny{$\pm$0.02}   &80.62\tiny{$\pm$0.02}   &95.36\tiny{$\pm$0.01}  \\
    DINO ViT-B/16     & 77.07\tiny{$\pm$0.20} &92.52\tiny{$\pm$0.09}   &\textbf{80.83\tiny{$\pm$0.16} }  &\textbf{95.53\tiny{$\pm$0.05}}   &\textbf{82.78\tiny{$\pm$0.17}}  &\textbf{96.42\tiny{$\pm$0.05}}  \\
    % DINO ViT-S/16    &  &   &   &   &   &  \\
    DINOv2 ViT-B/14  &\textbf{77.37\tiny{$\pm$0.25}}  &\textbf{92.91\tiny{$\pm$0.27}}   &79.55\tiny{$\pm$0.08}   &94.41\tiny{$\pm$0.03}   &81.88\tiny{$\pm$0.50}   &95.34\tiny{$\pm$0.20}  \\
    % DINOv2 ViT-S/14    &  &   &   &   &   &  \\
    % MedSAM ViT-B/16     &  &   &   &   &   &  \\
    %  BiomedGPT   &  &   &   &   &   &  \\
           \hline
  \end{tabular}}
  }
\end{table}
\end{savenotes}

\clearpage
\section{Plots of Performance Comparison on DermaMNIST with Different Image Sizes and Image Resizing Strategies}\label{appendix:results_plot_resize}

\begin{figure}[h]
 % Caption and label go in the first argument and the figure contents
 % go in the second argument
\floatconts
  {fig:ft_scale_sizes}
  {\caption{Performance comparison of foundation models on DermaMNIST with different image sizes, when using end-to-end fine-tuning and scaling for image resizing.}}
  {\includegraphics[width=0.8\linewidth]{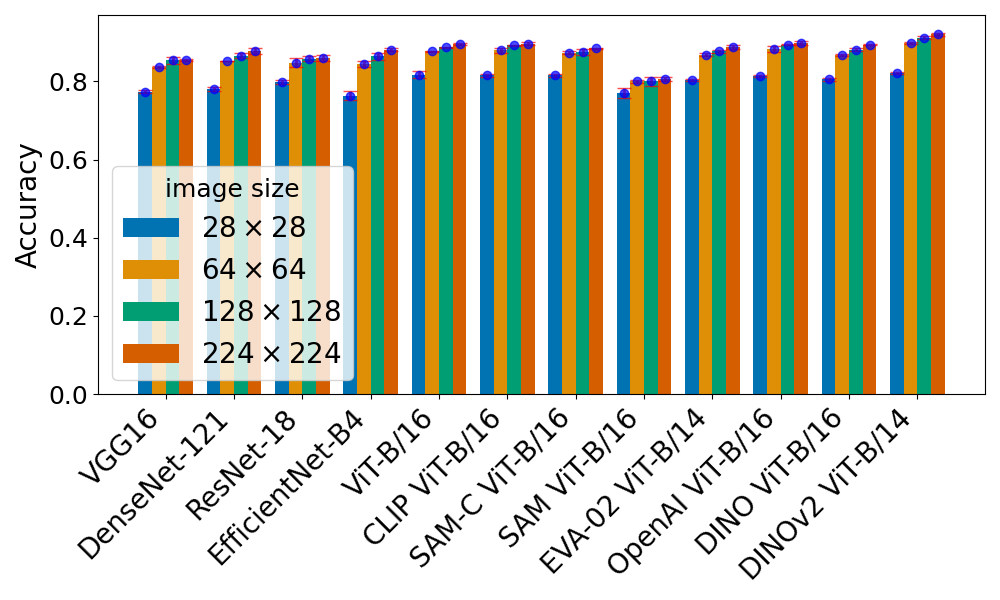}%
}
\end{figure}

\begin{figure}[h]
 % Caption and label go in the first argument and the figure contents
 % go in the second argument
\floatconts
  {fig:ft_pad_sizes}
  {\caption{Performance comparison of foundation models on DermaMNIST with different image sizes, when using end-to-end fine-tuning and zero-padding for image resizing.}}
  {\includegraphics[width=0.8\linewidth]{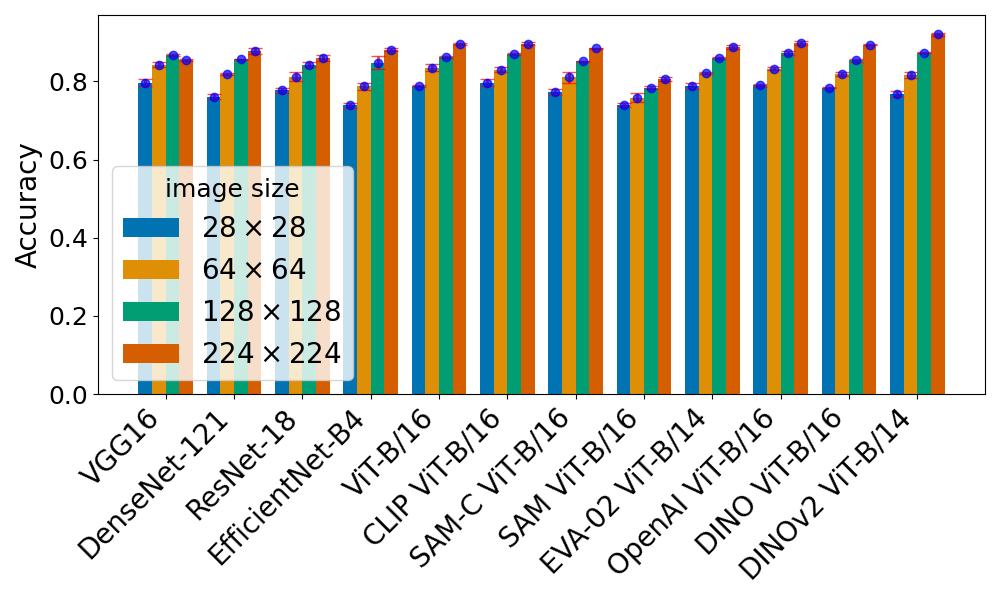}%
}
\end{figure}

\begin{figure}[h]
 % Caption and label go in the first argument and the figure contents
 % go in the second argument
\floatconts
  {fig:fz_scale_sizes}
  {\caption{Performance comparison of foundation models on DermaMNIST with different image sizes, when using linear probing and scaling for image resizing.}}
  {\includegraphics[width=0.8\linewidth]{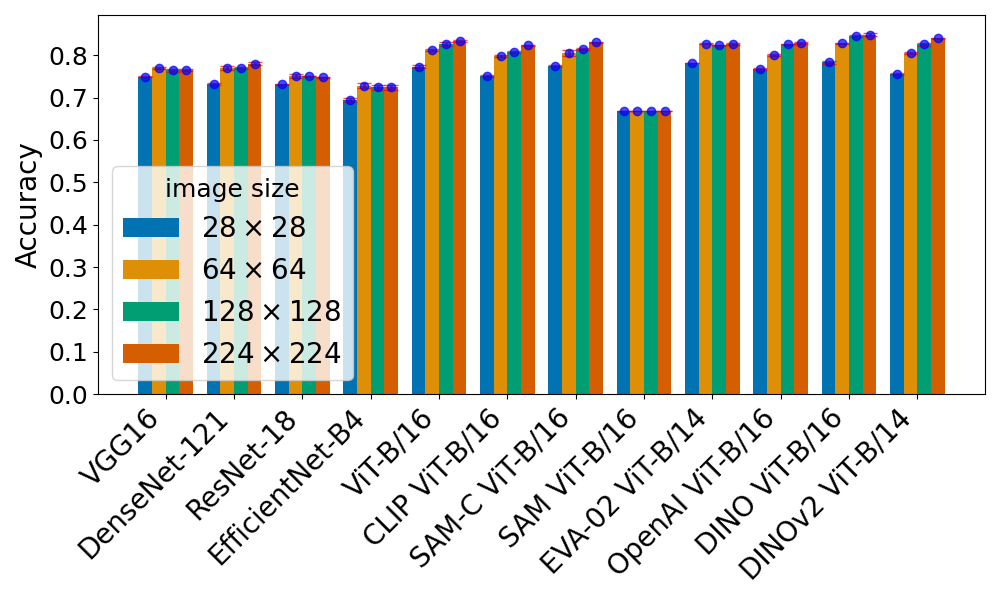}%
}
\end{figure}

\begin{figure}[h]
 % Caption and label go in the first argument and the figure contents
 % go in the second argument
\floatconts
  {fig:fz_pad_sizes}
  {\caption{Performance comparison of foundation models on DermaMNIST with different image sizes, when using linear probing and zero-padding for image resizing.}}
  {\includegraphics[width=0.8\linewidth]{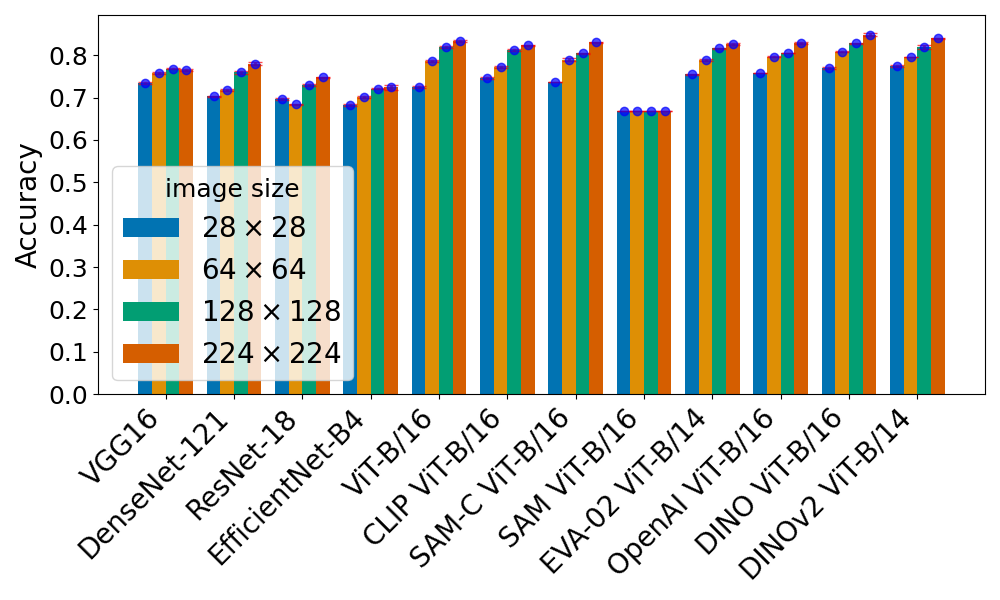}%
}
\end{figure}

\begin{figure}[h]
 % Caption and label go in the first argument and the figure contents
 % go in the second argument
\floatconts
  {fig:ft_28}
  {\caption{Performance comparison for zero-padding and scaling on DermaMNIST with an image size of $28\times 28$, when using end-to-end fine-tuning.}}
  {\includegraphics[width=0.8\linewidth]{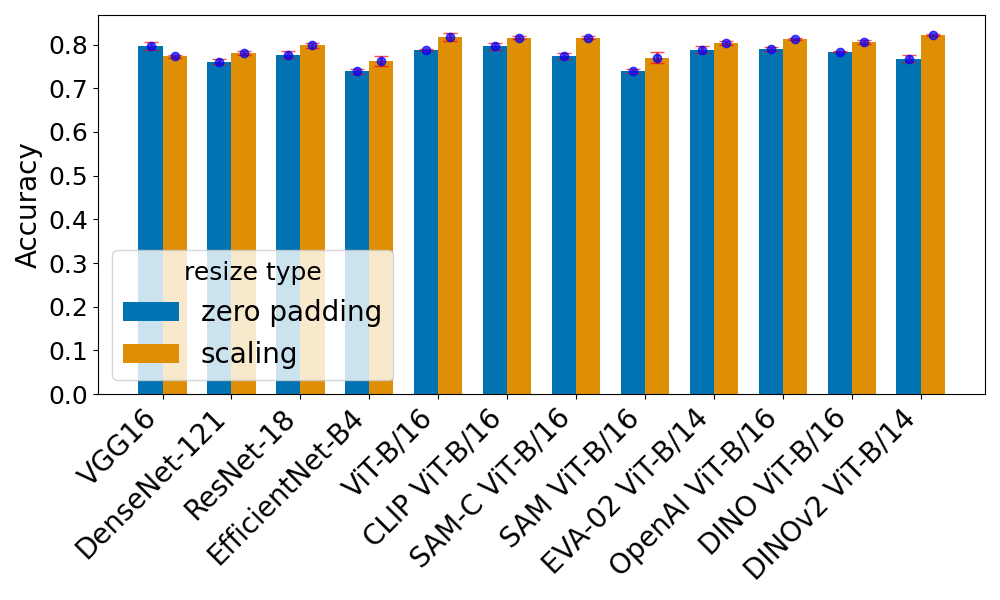}%
}
\end{figure}

\begin{figure}[h]
 % Caption and label go in the first argument and the figure contents
 % go in the second argument
\floatconts
  {fig:ft_64}
  {\caption{Performance comparison for zero-padding and scaling on DermaMNIST with an image size of $64\times 64$, when using end-to-end fine-tuning.}}
  {\includegraphics[width=0.8\linewidth]{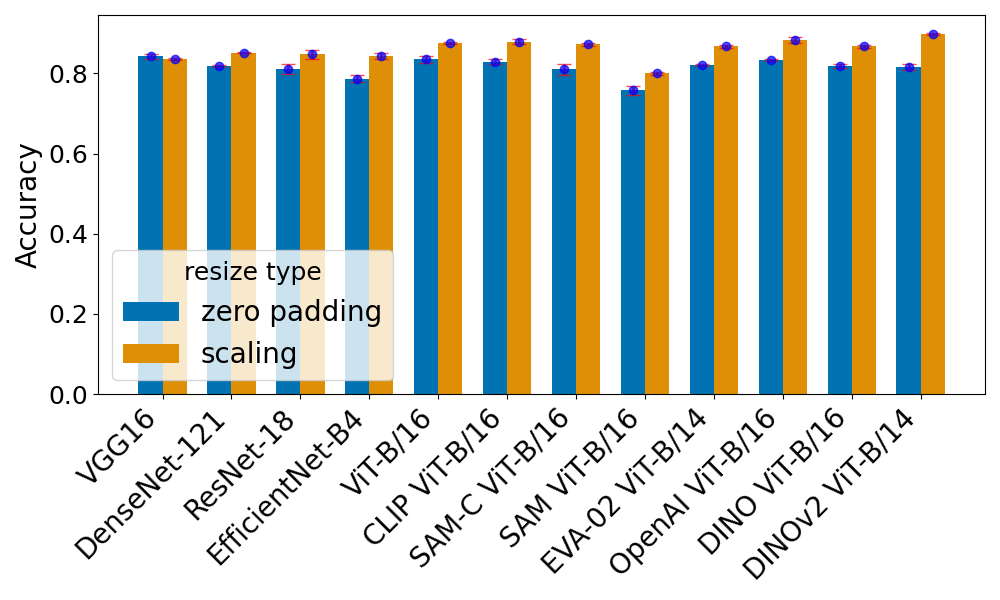}%
}
\end{figure}

\begin{figure}[h]
 % Caption and label go in the first argument and the figure contents
 % go in the second argument
\floatconts
  {fig:ft_128}
  {\caption{Performance comparison for zero-padding and scaling on DermaMNIST with an image size of $128\times 128$, when using end-to-end fine-tuning.}}
  {\includegraphics[width=0.8\linewidth]{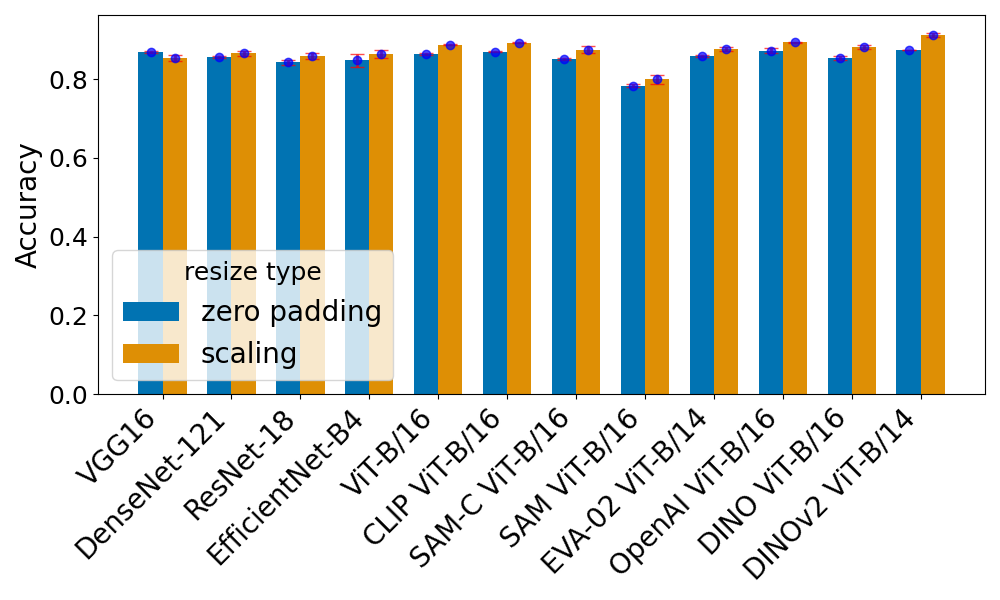}%
}
\end{figure}

\begin{figure}[h]
 % Caption and label go in the first argument and the figure contents
 % go in the second argument
\floatconts
  {fig:fz_28}
  {\caption{Performance comparison for zero-padding and scaling on DermaMNIST with an image size of $28\times 28$, when using linear probing.}}
  {\includegraphics[width=0.8\linewidth]{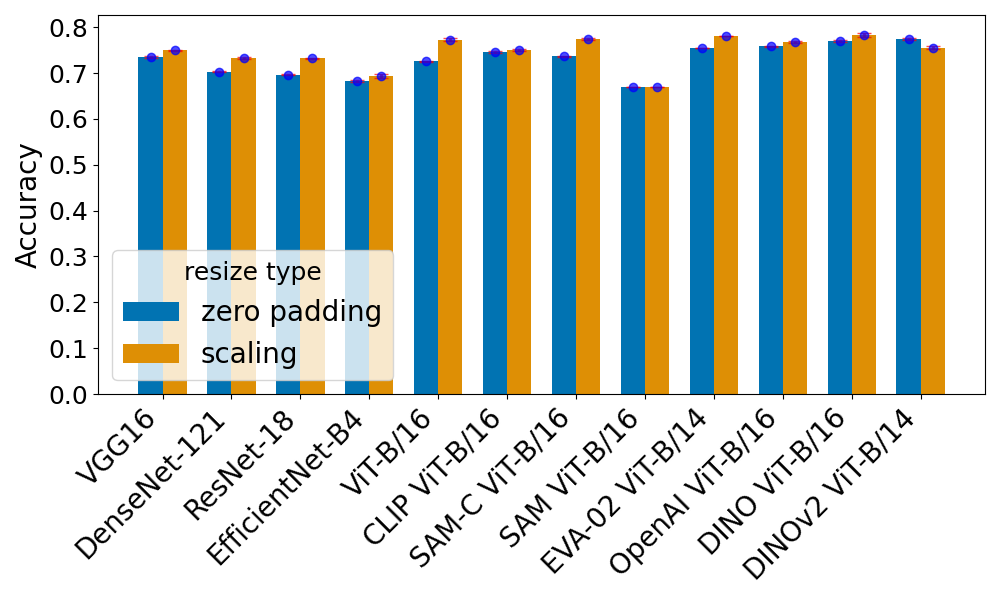}%
}
\end{figure}

\begin{figure}[h]
 % Caption and label go in the first argument and the figure contents
 % go in the second argument
\floatconts
  {fig:fz_64}
  {\caption{Performance comparison for zero-padding and scaling on DermaMNIST with an image size of $64\times 64$, when using linear probing.}}
  {\includegraphics[width=0.8\linewidth]{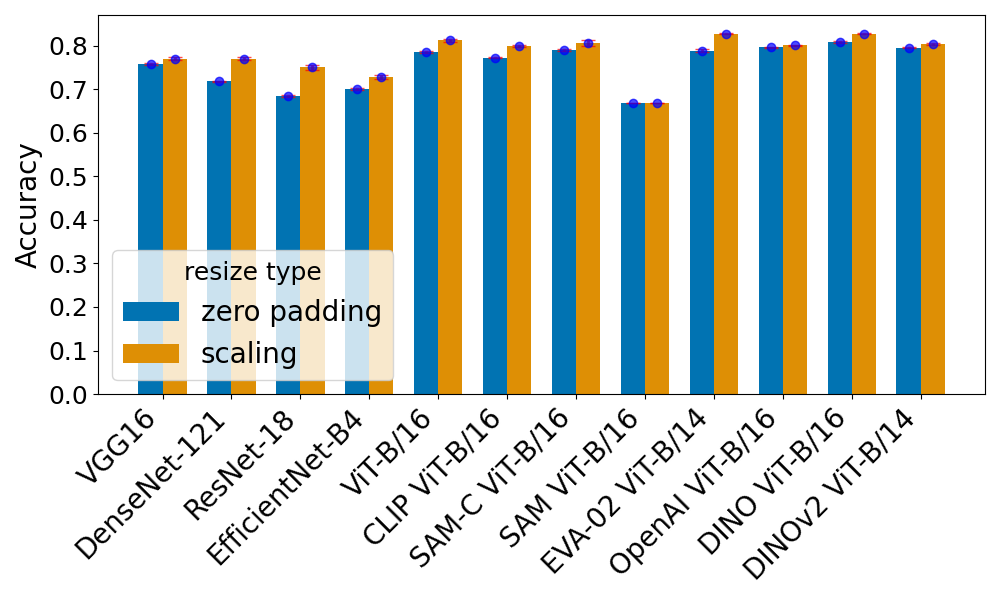}%
}
\end{figure}

\begin{figure}[h]
 % Caption and label go in the first argument and the figure contents
 % go in the second argument
\floatconts
  {fig:fz_128}
  {\caption{Performance comparison for zero-padding and scaling on DermaMNIST with an image size of $128\times 128$, when using linear probing.}}
  {\includegraphics[width=0.8\linewidth]{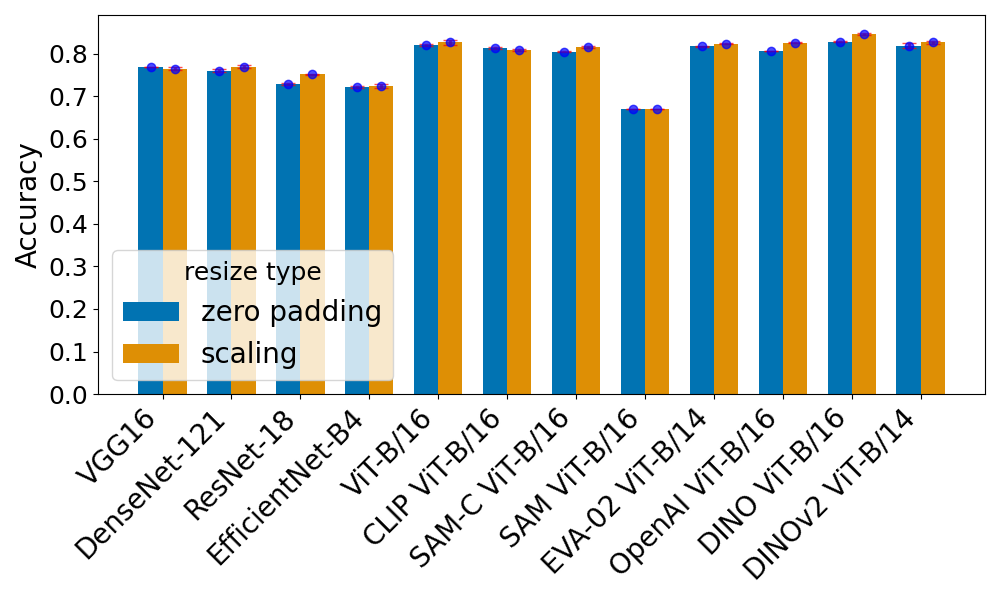}%
}
\end{figure}

% \begin{figure}[h]
%  % Caption and label go in the first argument and the figure contents
%  % go in the second argument
% \floatconts
%   {fig:few2}
%   {\caption{Accuracy of VGG16 on DermaMNIST when training model with a different number of data for each class.}}
%   {
% \includegraphics[width=0.8\linewidth]{Figs/Fewshot_perMethod_Derma/fewshot_vgg16_Accuracy.png}%
% }
% \end{figure}

\clearpage
\section{Plots of Performance Changes with Various Numbers of DermaMNIST Training Data} \label{appendix:plots_few_shot}

\begin{figure}[h]
 % Caption and label go in the first argument and the figure contents
 % go in the second argument
\floatconts
  {fig:few3}
  {\caption{Accuracy of DenseNet-121 on DermaMNIST when training model with a different number of data for each class.}}
  {
\includegraphics[width=0.8\linewidth]{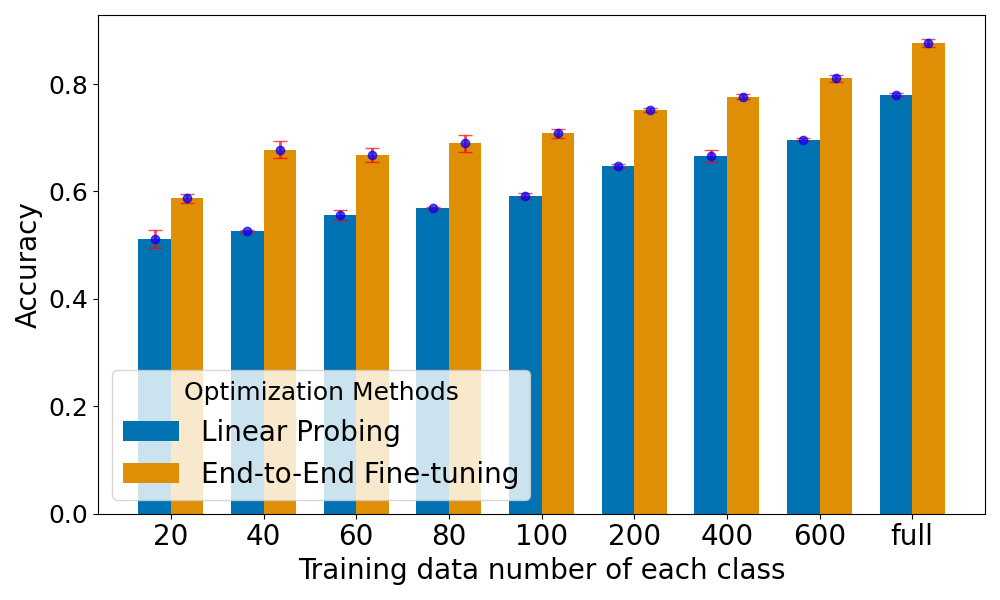}%
}
\end{figure}

\begin{figure}[h]
 % Caption and label go in the first argument and the figure contents
 % go in the second argument
\floatconts
  {fig:few4}
  {\caption{Accuracy of ResNet-18 on DermaMNIST when training model with a different number of data for each class.}}
  {
\includegraphics[width=0.8\linewidth]{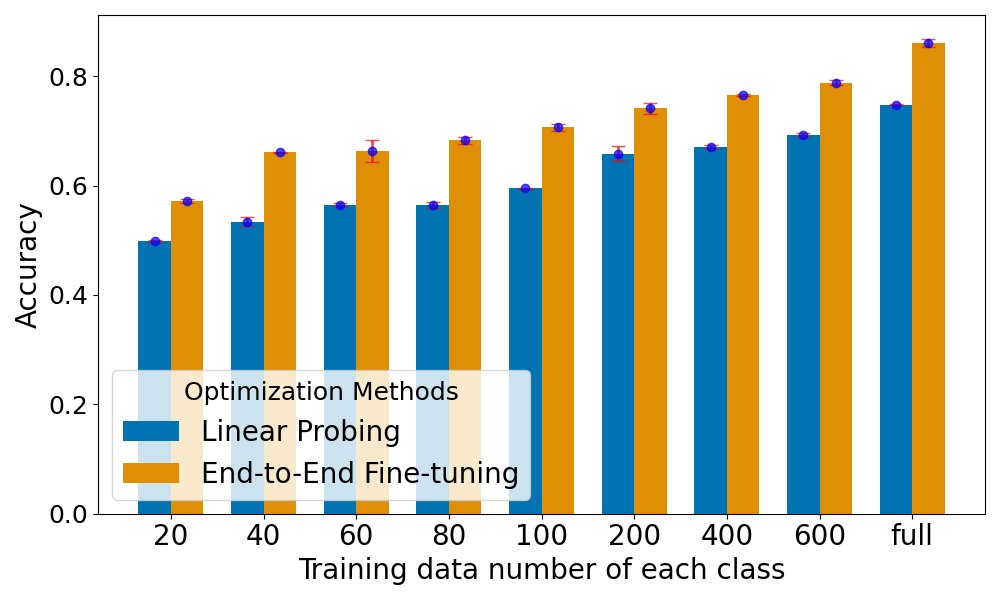}%
}
\end{figure}

\begin{figure}[h]
 % Caption and label go in the first argument and the figure contents
 % go in the second argument
\floatconts
  {fig:few5}
  {\caption{Accuracy of EfficientNet-B4 on DermaMNIST when training model with a different number of data for each class.}}
  {
\includegraphics[width=0.8\linewidth]{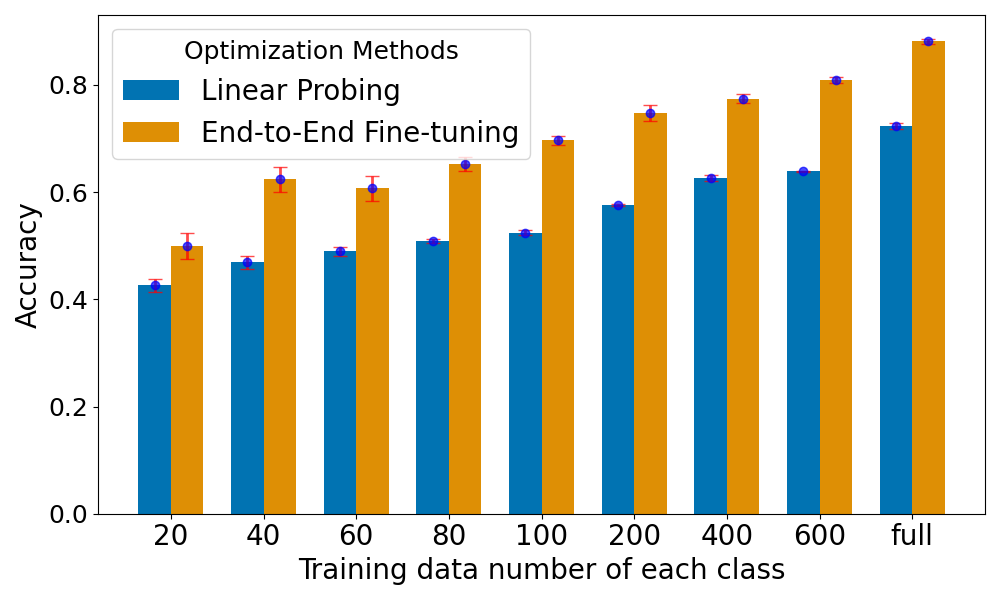}%
}
\end{figure}

\begin{figure}[h]
 % Caption and label go in the first argument and the figure contents
 % go in the second argument
\floatconts
  {fig:few6}
  {\caption{Accuracy of ViT-B/16 on DermaMNIST when training model with a different number of data for each class.}}
  {
\includegraphics[width=0.8\linewidth]{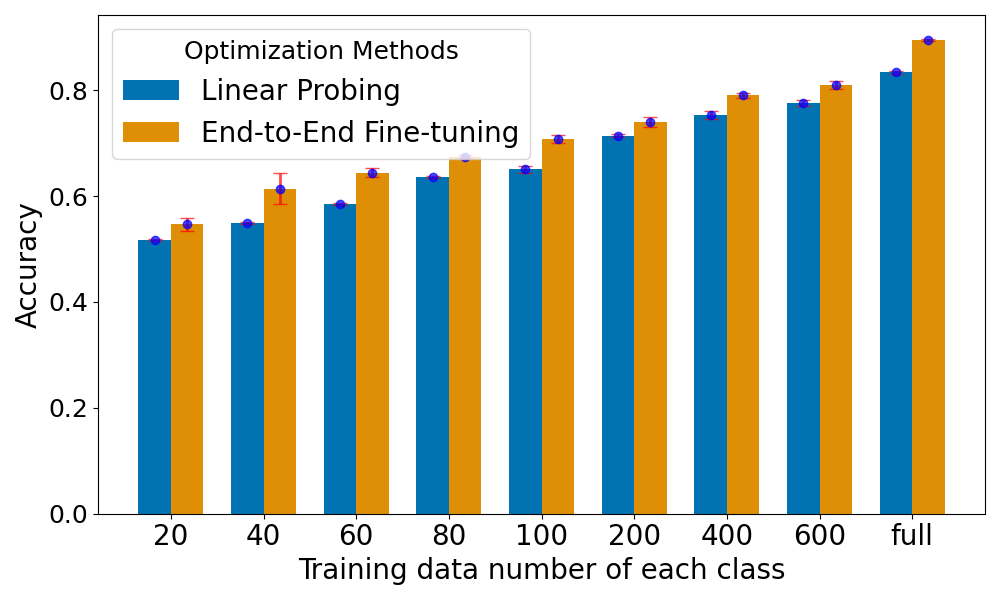}%
}
\end{figure}

\begin{figure}[h]
 % Caption and label go in the first argument and the figure contents
 % go in the second argument
\floatconts
  {fig:few7}
  {\caption{Accuracy of CLIP ViT-B/16 on DermaMNIST when training model with a different number of data for each class.}}
  {
\includegraphics[width=0.8\linewidth]{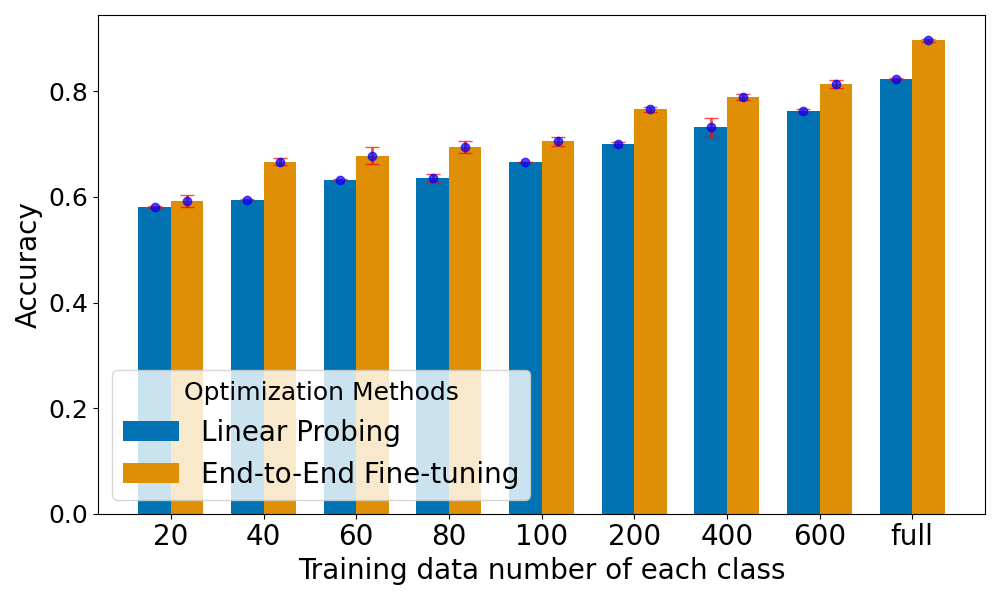}%
}
\end{figure}

\begin{figure}[h]
 % Caption and label go in the first argument and the figure contents
 % go in the second argument
\floatconts
  {fig:few8}
  {\caption{Accuracy of SAM-C ViT-B/16 on DermaMNIST when training model with a different number of data for each class.}}
  {
\includegraphics[width=0.8\linewidth]{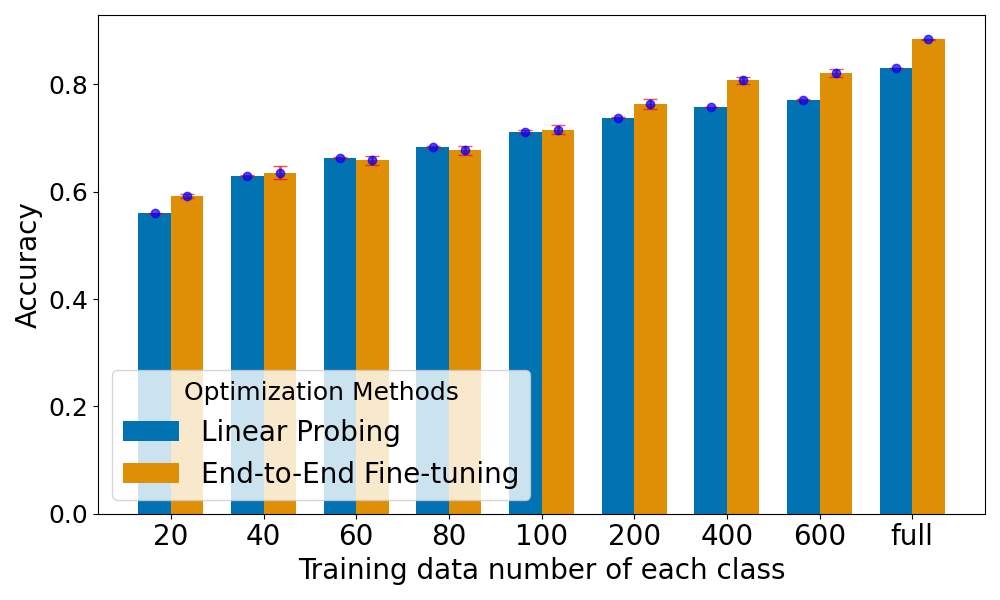}%
}
\end{figure}

\begin{figure}[h]
 % Caption and label go in the first argument and the figure contents
 % go in the second argument
\floatconts
  {fig:few9}
  {\caption{Accuracy of SAM ViT-B/16 on DermaMNIST when training model with a different number of data for each class.}}
  {
\includegraphics[width=0.8\linewidth]{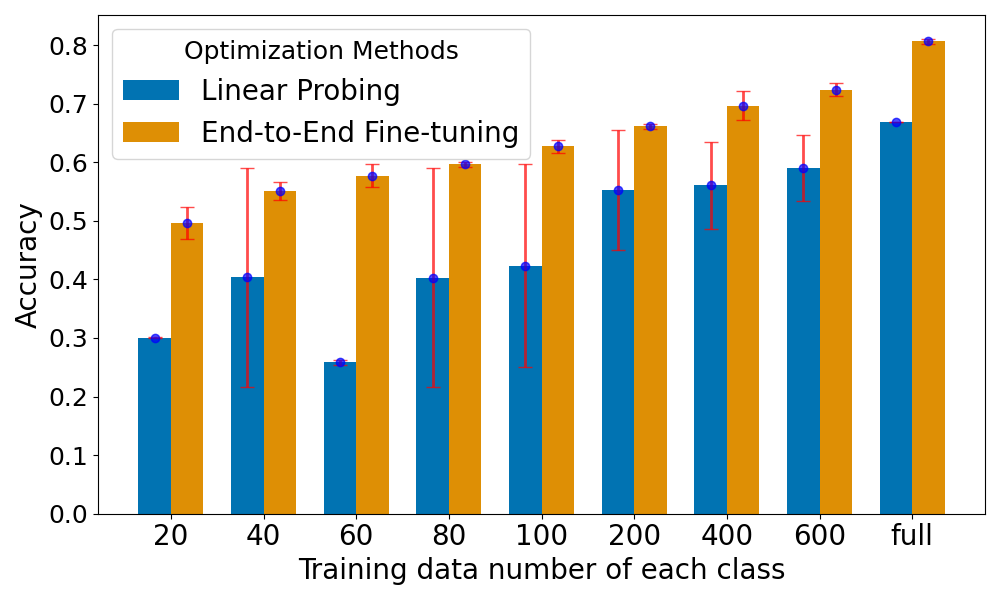}%
}
\end{figure}

\begin{figure}[h]
 % Caption and label go in the first argument and the figure contents
 % go in the second argument
\floatconts
  {fig:few10}
  {\caption{Accuracy of EVA-02 ViT-B/14 on DermaMNIST when training model with a different number of data for each class.}}
  {
\includegraphics[width=0.8\linewidth]{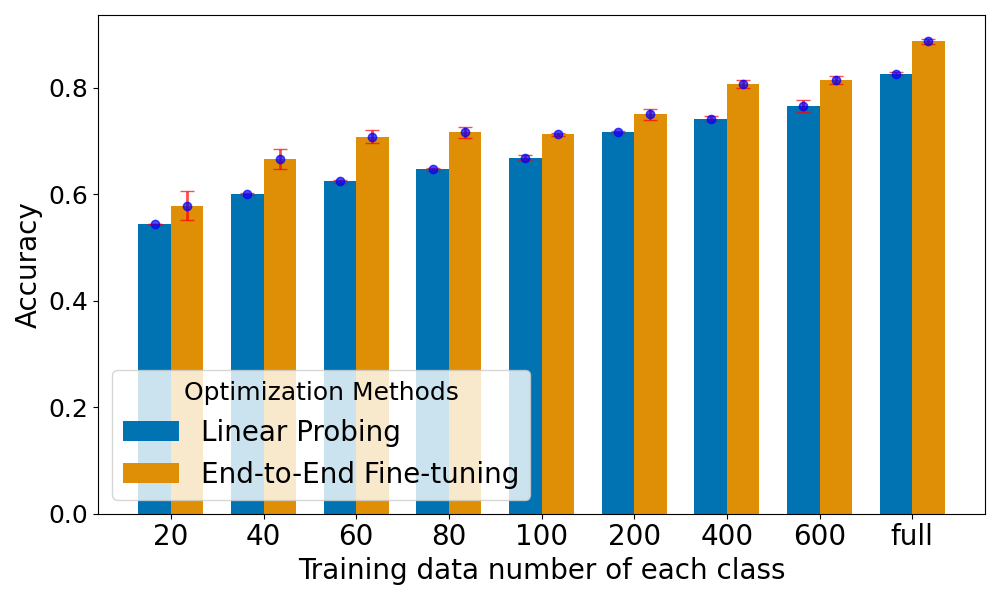}%
}
\end{figure}

\begin{figure}[h]
 % Caption and label go in the first argument and the figure contents
 % go in the second argument
\floatconts
  {fig:few11}
  {\caption{Accuracy of DINO OpenAI ViT-B/16 on DermaMNIST when training model with a different number of data for each class.}}
  {
\includegraphics[width=0.8\linewidth]{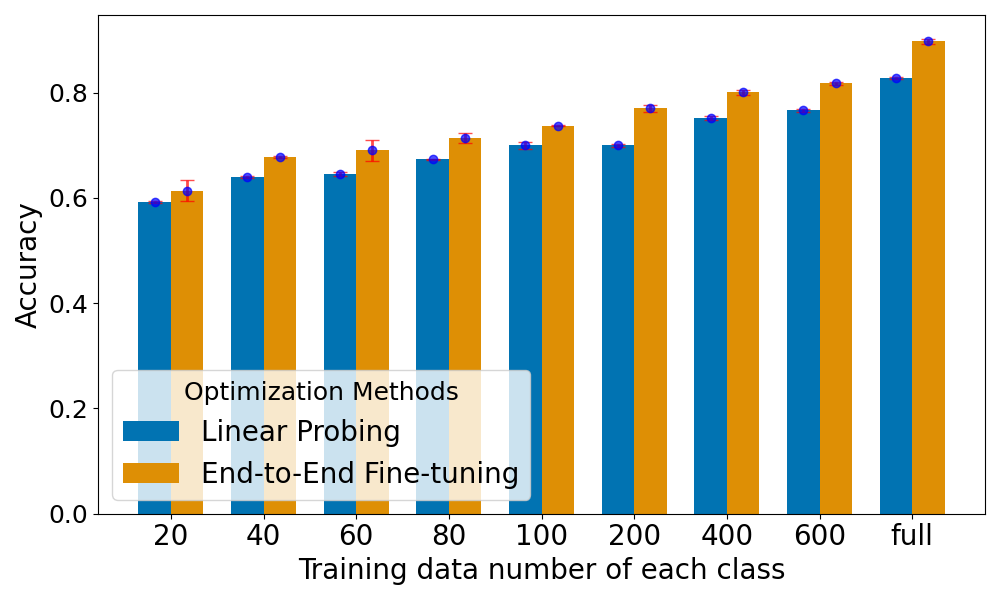}%
}
\end{figure}

\begin{figure}[h]
 % Caption and label go in the first argument and the figure contents
 % go in the second argument
\floatconts
  {fig:few12}
  {\caption{Accuracy of DINOv2 ViT-B/14 on DermaMNIST when training model with a different number of data for each class.}}
  {
\includegraphics[width=0.8\linewidth]{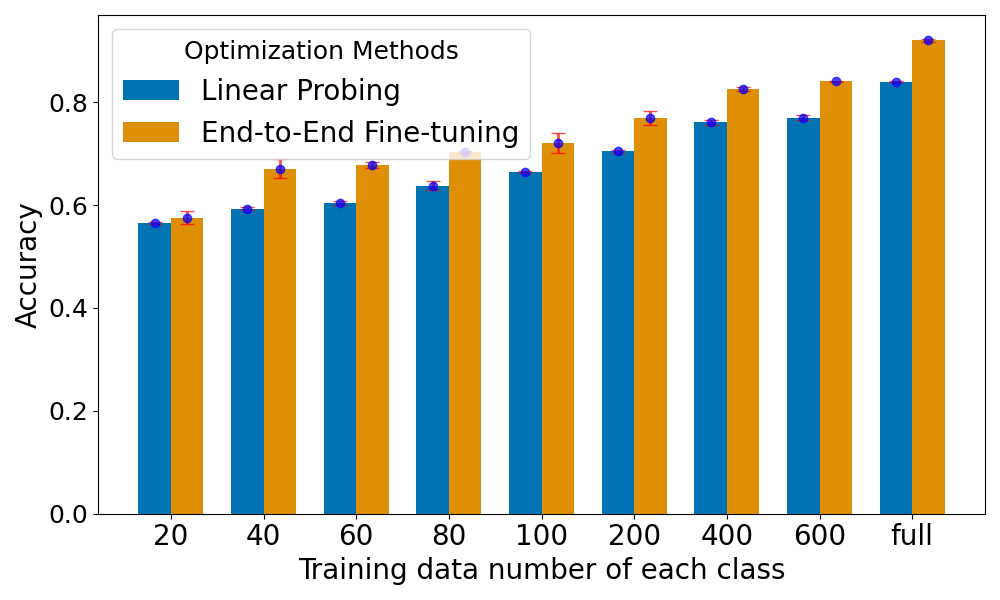}%
}
\end{figure}

\end{document}